\documentclass[a4paper,oneside]{article}

\usepackage{amsmath}
\usepackage{amsfonts}

\addtocounter{equation}{1}
\newcommand{\dbar}{d\!\!\!\lower-0.6ex\hbox{${-}$\!\!}}
\newcommand{\e}{\varepsilon}

\newcommand{\symsum}{\hbox{$\sum
     $\kern-1em\lower-0.3ex\hbox{$\scriptscriptstyle{\bigcirc}$}}\,}

\newcommand{\ott}{\lower-0.4ex\hbox{${\scriptscriptstyle{\otimes}}$}}
\newcommand{\btt}{\lower-0.2ex\hbox{${\scriptscriptstyle{\bullet}}$}}
\newcommand{\ctt}{\lower-0.2ex\hbox{${\scriptscriptstyle{\circ}}$}}
\newcommand{\dtt}{\lower-0.2ex\hbox{${\scriptscriptstyle{\odot}}$}}
\newcommand{\ett}{\lower-0.2ex\hbox{${\scriptscriptstyle{\lozenge}}$}}
\newtheorem{thm}{Theorem}[section]
\newtheorem{prop}{Proposition}[section]
\newtheorem{definition}{Definition}[section]

\numberwithin{equation}{section}

\begin{document}

\begin{title}
{\bf Deformation  Expression for Elements of Algebra}
\end{title}
\author{Hideki Omori 
   \thanks{Department of Mathematics, Faculty of Science and Technology, 
      Tokyo University of Science, Noda, Chiba, 278-8510, Japan,
    email: omori@ma.noda.sut.ac.jp}
    \\Tokyo University of Science
\and Yoshiaki Maeda 
   \thanks{Department of Mathematics, Faculty of Science and Technology,
     Keio University, Hiyoshi, Yokohama, 223-8522, Japan,
        email: maeda@math.keio.ac.jp}
   \\ Keio University
\and Naoya Miyazaki 
   \thanks{Department of Mathematics, Faculty of Economics,
     Keio University, Hiyoshi, Yokohama, 223-8522, Japan,
 email: Naoya.Miyazaki@math.yokohama-cu.ac.jp}
   \\Keio University
\and Akira Yoshioka 
   \thanks{Department of Mathematics, Faculty of Engineering, 
        Tokyo University of Science, 
            Kagurazaka, Shinjyuku-ku, Tokyo 162-8601, Japan,
            email: yoshioka@rs.kagu.sut.ac.jp}
  \\Tokyo University of Science 
}
\maketitle

\section{Introduction}

The purpose of this paper is to give a notion of 
deformation of expressions for elements of algebra. 

Deformation quantization (cf.[BF]) deforms the commutative 
world to a non-commutative world. However, this 
involves deformation of expression of elements of algebras     
even from a commutative world to another commutative world. 
This is indeed a deformation of expressions for  
elements of algebra. 

\medskip

\section{Definition of $*$-functions and intertwiners}
Let ${\mathbb C}[w]$ be the space of polynomials of 
one variable $w$. For a complex parameter $\tau$, 
we define a new product on this space 
\begin{equation}
 \label{eq:prodgen}
 f{*}_{\tau}g=\sum_{k\geq 0}
\frac{\tau^k}{2^kk!}\partial^k_{w}f\partial^k_{w}g  
\quad(\,\,=fe^{\frac{\tau}{2}
\overleftarrow{\partial_{{w}}}\overrightarrow{\partial_{{w}}}}g). 
\end{equation} 
We see easily that $*_{\tau}$ makes 
${\mathbb C}[w]$ a commutative associative algebra, 
which we denote by 
$({\mathbb C}[w], *_{\tau})$. If $\tau{=}0$, then 
$({\mathbb C}[w], *_{0})$ is the usual polynomial algebra, and 
$\tau\in {\mathbb C}$ is called a deformation parameter.
What is deformed is not the algebraic structure, 
but the expression of elements. 

\subsection{Intertwiners and infinitesimal intertwiners}

It is not hard to verify that the mapping  
\begin{equation}
 \label{eq:intwnner}
e^{\frac{\tau}{4}\partial^2_{w}}: ({\mathbb C}[w],{*}_{0})\to
({\mathbb C}[w],{*}_{\tau})   
\end{equation}
gives an algebra isomorphism. That is, 
$e^{\frac{\tau}{4}\partial^2_{w}}$ has the inverse mapping   
$e^{-\frac{\tau}{4}\partial^2_{w}}$ and 
$ e^{\frac{\tau}{4}\partial^2_{w}}(f{*}_0g){=}
(e^{\frac{\tau}{4}\partial^2_{w}}f){*}_{\tau}
(e^{\frac{\tau}{4}\partial^2_{w}}g)$
 holds.
The isomorphism 
$I_0^{\tau}{=}e^{\frac{\tau}{4}\partial^2_{w}}$ 
is called the {\bf intertwiner}.  
Defining  
$I_{\tau}^{\tau'}=I_0^{\tau'}(I_0^{\tau})^{-1}$ 
gives the intertwiner from $({\mathbb C}[w],{*}_{\tau})$ 
onto $({\mathbb C}[w],{*}_{\tau'})$.
Its differential 
$dI_{\tau}{=}I_{\tau}^{\tau{+}d\tau}{=}
\frac{d}{d\tau'}I_{\tau}^{\tau'}\Big|_{\tau'=\tau}{=}\frac{1}{4}\partial^2_{w}$
is called the {\bf infinitesimal intertwiner}.

\medskip 
Defining $w_{*\tau}^n$ by $I_0^{\tau}w^n$ we get    
\begin{equation}\label{Pn00}
w_{*\tau}^n{=}
P_n(w,\tau){=}
\sum_{k\leq [n/2]}\frac{n!}{4^kk!(n{-}2k)!}\tau^kw^{n{-}2k}.
\end{equation}
%
%
\medskip
Let $H\!ol({\mathbb C})$ be the space of all entire functions on 
${\mathbb C}$ with the topology of uniform convergence 
on each compact domain. 
$H\!ol({\mathbb C})$ is known to be a Fr{\'e}chet space 
defined by a countable family of seminorms.
It is easy to see that the product $*_{\tau}$ extends naturally for  
$f, g\in H\!ol({\mathbb C})$ if either $f$ or $g$ is a polynomial. 
By the inductive limit topology 
${\mathbb C}[w]$ is a complete topological algebra 
with uncountable basis of neighborhods of $0$. 
We easily see the following: 
\begin{thm}\label{fund00pol}
For a polynomial $p(w)$, the multiplication 
$p(w){*_\tau}$  is a continuous linear mapping 
of $H\!ol({\mathbb C})$ into itself. 
By polynomial approximations, the associativity 
$f{*_\tau}(g{*_\tau}h)=(f{*_\tau}g){*_\tau}h$ holds 
if two of $f,g,h$ are polynomials.    
$H\!ol({\mathbb C})$ is a topological 
${\mathbb C}[w]$ bi-module. 
\end{thm}

\subsection{$*$-exponential functions and 
$\tau$-expressions}\label{111}
We now study the deformation of the exponential function $e^{aw}$.
Although the ordinary exponential function $e^{aw}$ is not 
a polynomial, the intertwiner $I_0^{\tau}$ given by 
 \eqref{eq:intwnner} extends to give 
\begin{equation}
  \label{eq:expdef}
 I_0^{\tau}(e^{2aw}){=}e^{2aw{+}a^2\tau}{=}
e^{a^2\tau}e^{2aw}, \quad \tau\in{\mathbb C}
\end{equation}
Using Taylor expansion, we get  
\begin{equation}
\label{eq:prodexp}
e^{2aw}{*_{\tau}}e^{2bw}{=}
e^{2(a{+}b)w{+}2ab\tau},\quad
e^{2aw}{*_{\tau}}f(w)=e^{2aw}f(w{+}a\tau)
\end{equation}
for every $f{\in}H\!ol({\mathbb C})$. 
We have also the associativity 
$e^{2aw}{*_\tau}(e^{2bw}{*_\tau}f(w))
{=}(e^{2a\tau}{*_\tau}e^{2bw}){*_\tau}f(w)$ 
for every $f{\in}H\!ol(\mathbb C)$.   
Computation via intertwiners gives     
$I^{\tau'}_{\tau}(e^{\frac{1}{4}s^2\tau}e^{s{w}})
{=}e^{\frac{1}{4}s^2\tau'}e^{s{w}}.$
Denoting $e_*^{s{w}}$ by the family  
$\{e^{\frac{1}{4}s^2\tau}e^{s{w}}; \tau\in{\mathbb C}\}$ we call this 
the $*$-{\bf exponential function}. 
  
Associated with polynomials and exponential functions $f(w)$, 
we construct a family of functions  
\begin{equation}\label{eq2.6}
\{f_{{\tau}}(w); \tau{\in}{\mathbb C}\}, \quad f_{\tau}{=}I_0^{\tau}(f(w)),
\end{equation}
which is denoted by $f_*(w)$. We view $f_*(w)$ as an {\it element} of 
the abstract algebra. Given $f$ we refer to the object \eqref{eq2.6} as a 
$*$-{\bf function}. By using the notation ${:}\btt{:}_{\tau}$ we
denote as  
\begin{equation*}
  \label{eq:tempexp}
{:}f_*(w){:}_\tau{=}f_\tau(w) 
\end{equation*}
${:}f_*{:}_{\tau}$ is 
viewed as the $\tau$-{\bf expression} of $f_*$. 
Then we have  
${:}e_*^{sw}{:}_{\tau}=e^{\frac{1}{4}s^2\tau}e^{s{w}}$ 
and we call the r.h.s the $\tau$-expression of $e_*^{s{w}}$. 
The product formula \eqref{eq:prodgen} gives 
the exponential law
\begin{equation}
  \label{eq:explaw1}
{:}e_*^{s{w}}{:}_{\tau}{*_{\tau}}{:}e_*^{t{w}}{:}_{\tau}=
{:}e_*^{(s{+}t){w}}{:}_{\tau},
\quad \forall \tau\in {\mathbb C}.  
\end{equation}
Note that ${:}e_*^{t{w}}{:}_{\tau}$ is the solution for every 
$\tau$ of the differential equation  
$\frac{d}{dt}g(t)={w}{*_{\tau}}g(t)$  
with the initial condition $g(0){=}1$.
It is easy to see the exponential law $e_*^{tw}e^{s}=e_*^{tw{+}s}$ 
holds for the ordinary exponential function $e^{s}$.
The formula  
${:}e_*^{tw}{:}_\tau{=}\sum_n \frac{t^n}{n!}{:}w_*^n{:}_\tau$ also holds. 

For every $f\in H\!ol(\mathbb C)$, the formula \eqref{eq:prodexp} gives 
\begin{equation} 
 \label{eq:prodexp22}
{:}e_*^{2s{w}}{:}_\tau{*_\tau}f({w})
=e^{2s{w}{+}s^2\tau}f({w}{+}s\tau)   
\end{equation} 
Using this, we have several basic properties of $*$-exponential functions: 
\begin{prop}\label{ketugoprodexp22}  
The associativity 
$e_*^{rw}{*}(e_*^{sw}{*}f){=}e_*^{(r{+}s)w}{*}f{=}e_*^{rw}{*}(f{*}e_*^{sw})$ 
holds in every $\tau$-expression. 

If $f(w)\in
H\!ol({\mathbb C})$ satisfies
${:}e_*^{isw}{:}_{\tau}{*}_{\tau}f(w)=0$,  then $f(w)=0$. 
\end{prop}

As ${:}e_*^{2niw}{:}_{\tau}=e^{-n^2\tau}e^{2niw}$, 
if ${\rm{Re}}\tau{>}0$, then ${:}e_*^{2niw}{:}_{\tau}$ 
tends to $0$ very quickly. Using this we have 
\begin{prop}\label{entirelevel}
If a power series $\sum_{n=0}^{\infty}a_nz^n$ has a positive radius of
convergence, then  
${:}e_*^{\ell iw}{:}_{\tau}
{*}_{\tau}\sum_{n=0}^{\infty}a_n{:}e_*^{niw}{:}_{\tau}$ 
is an entire function of $w$ for every $\ell\in {\mathbb Z}$. 
\end{prop}

On the other hand we note the following:
\begin{prop}\label{convzero}
If $\ell\geq 3$ and $\tau\not=0$, 
then the radius of convergence of the power series  
$\sum_{n=0}^\infty
\frac{t^n}{n!}{:}w_*^{n\ell}{:}_{\tau}$ 
in $t$ is $0$. That is, 
$e_*^{tw_*^{\ell}}$ can not be defined as a power series for 
$\ell\geq 3$.
\end{prop}

\subsection{Applications to generating functions} 

We note that exponential functions contribute to construct generating
functions. We show how $*$-exponential functions 
relates to generating functions.

\noindent
{\bf The generating function of Hermite polynomials} is 
given by $e^{\sqrt{2}tx{-}\frac{1}{2}t^2}
=\sum_{n=0}^\infty H_n(x)\frac{t^n}{n!}.$ 
This is the Taylor expansion formula of 
${:}e_*^{\sqrt{2}tw}{:}_{-1}$. 

Noting $e_*^{taw}=\sum\frac{t^n}{n!}(aw)_*^n$, and setting 
$e_*^{\sqrt{2}tw}{=}\sum_{n\geq 0}(\sqrt{2}w)_*^n\frac{t^n}{n!},$
we see $H_n(w)={:}(\sqrt{2}w)_*^n{:}_{-1}$. Hence it is easy to see
that $H_n(w)$ is a polynomial of degree $n$.
We define for every $\tau\in \mathbb C$ $*$-Hermite polynomials $H_n(w,*)$ by 
\begin{equation}\label{hermite}
e_*^{\sqrt{2}tw}{=}\sum_{n\geq 0}^\infty H_n(w,*)\frac{t^n}{n!},
\qquad (H_n(w,\tau){=}{:}H_n(w,*){:}_{\tau},\quad H_n(w,-1){=}H_n(w)).
\end{equation} 
Since 
$\frac{d}{dt}e_*^{\sqrt{2}tw}{=}\sqrt{2}w{*}e_*^{\sqrt{2}tw}$, 
we have 
$\frac{\tau}{\sqrt{2}}H'_n(w,\tau){+}\sqrt{2}wH_n(w,\tau)=H_{n{+}1}(w,\tau)$ 
where $H'_n(w,\tau){=}\frac{\partial}{\partial w}H(w,\tau)$.

\bigskip
The exponential law yields 
$\sum_{k+\ell=n}\frac{n!}{k!{\ell}!}H_k(w,*){*}H_{\ell}(w,*)=
H_n(w,*).$
On the other hand, taking 
$\frac{\partial}{\partial w}$ of both sides of 
\eqref{hermite} gives $\sqrt{2}nH_{n{-}1}(w,*)=H'_n(w,*).$
Differentiate again and use the above equality to get 
$$
{\tau}H''_n(w,\tau){+}2wH'_n(w,\tau){-}2nH_{n}(w,\tau)=0.
$$

By setting 
$\sqrt{2}tw+\frac{\tau}{2}t^2=
\frac{\tau}{2}(t{+}\frac{\sqrt{2}}{\tau}w)^2-\frac{1}{\tau}w^2$ 
the Hermite polynomial $H_n(w,*)$ is obtained via the following formula:
$$
H_n(w,\tau){=}
\frac{d^n}{dt^n}
e^{\frac{\tau}{2}(t{+}\frac{\sqrt{2}}{\tau}w)^2}\Big|_{t=0}e^{-\frac{1}{\tau}w^2}
=e^{-\frac{1}{\tau}w^2}(\frac{\tau}{\sqrt{2}})^n\frac{d^n}{dw^n}e^{\frac{1}{\tau}{w^2}}
$$
The orthogonality of $\{H_n(w,\tau)\}_n$ is shown  under
the condition ${\rm{Re}}\tau<0$ as follows:
$$
\int_{\mathbb R}e^{\frac{1}{\tau}w^2}H_n(w,\tau)H_m(w,\tau)dw=
\int_{\mathbb R}(\frac{\tau}{\sqrt{2}})^n\frac{d^n}{dw^n}
e^{\frac{1}{\tau}{w^2}} H_m(w,\tau)dw.
$$
If $n\not= m$, one may suppose $n>m$ without loss of generality. 
Hence this vanishes by the integration by parts $n$ times.
For the case $n=m$,  we set  
${:}e_*^{\sqrt{2}tw}{:}_{\tau}{=}e^{\frac{\tau}{2}t^2{+}\sqrt{2}tw}
=\sum_{n=0}^\infty H_n(w,\tau)\frac{t^n}{n!}$. 
Hence we see 
$$
\frac{1}{n!}H_n(w,\tau)=\sum_{p=0}^{[n/2]}\frac{\sqrt{2}^n\tau^p}{p!(n-2p)!4^p}w^{n-2p},\quad 
\frac{d^n}{dw^n}H_n(w,\tau)=\sqrt{2}^nn!.
$$

It follows 
$$
\int_{\mathbb R}e^{\frac{1}{\tau}w^2}H_n(w,\tau)H_n(w,\tau)dw=
n!(-\tau)^n\int_{\mathbb R}e^{\frac{1}{\tau}w^2}dw=
n!(-\tau)^n\sqrt{-\tau}\sqrt{\pi}. 
$$

\noindent
{\bf The generating function of Bessel functions} $J_n(z)$  is known to be 
$e^{iz\sin s}= \sum_{n=-\infty}^{\infty}J_n(z)e^{ins}.$ 
Keeping this in mind, we define $*$-Bessel functions by 
$$
e_*^{\frac{1}{2}(e^{is}{-}e^{-is})aw}= \sum_{n=-\infty}^{\infty}J_n(aw,*)e^{ins}, \quad 
{:}J_n(aw,*){:}_{\tau}=J_n(aw,\tau),\quad a\in{\mathbb C}.
$$
Replacing $s$ by $s{+}\frac{\pi}{2}$ gives 
$e_*^{\frac{i}{2}(e^{is}{+}e^{-is})aw}=\sum_{n=-\infty}^{\infty}i^nJ_n(aw,*)e^{ins}$
and basic symmetric properties hold: First we see 
$J_n(aw,*)=(-1)^nJ_{-n}(aw,*)$. Replacing $w$ by $-w$ in the first
equality gives $J_n(-aw,*)=J_{-n}(aw,*)$.    
Since 
$$
{:}e_*^{\frac{1}{2}(e^{is}{-}e^{-is})aw}{:}_{\tau}=
e^{\frac{a^2}{16}\tau(e^{is}{-}e^{-is})^2}e^{\frac{1}{2}(e^{is}{-}e^{-is})aw}
=e^{\frac{a^2}{8}\tau}
e^{-\frac{a^2}{16}\tau(e^{2is}{+}e^{-2is})}e^{\frac{1}{2}(e^{is}{-}e^{-is})aw},
$$
$J_n(aw,\tau)$ and $J_n(aw)$ are related by 
$$
\sum_{n=-\infty}^{\infty}J_n(aw,\tau)e^{ins}=
e^{\frac{a^2}{8}\tau}
e^{-\frac{a^2}{16}\tau(e^{2is}{+}e^{-2is})}
\sum_{n=-\infty}^{\infty}J_n(aw)e^{ins}.
$$
Setting $s=0$, we see in particular 
$1=\sum_{n=-\infty}^{\infty}J_n(aw,\tau)=\sum_{n=-\infty}^{\infty}J_n(aw).$
The exponential law of l.h.s. of the defining equality gives that 
$$
e_*^{\frac{1}{2}(e^{is}{-}e^{-is})aw}{*}e_*^{\frac{1}{2}(e^{is}{-}e^{-is})bw}
=e_*^{\frac{1}{2}(e^{is}{-}e^{-is})(a{+}b)w}=
\sum_nJ_n(aw{+}bw,*)e^{nis}.
$$
$$
J_n(aw{+}bw,*)=\sum_{m=-\infty}^{\infty}J_m(aw,*){*}J_{n{-}m}(bw). 
$$

If $a^2{+}b^2=1$, then  
$$
e_*^{\frac{1}{2}(e^{is}{-}e^{-is})aw}{*}e_*^{\frac{i}{2}(e^{is}{+}e^{-is})bw}
=e_*^{\frac{1}{2}((a{+}ib)e^{is}{-}(a{-}ib)e^{-is})w}=
\sum_nJ_n(w,*)(a{+}ib)^ne^{nis}.
$$
$$
\sum_{k=-\infty}^{\infty}J_k(aw,*)e^{iks}{*}\sum_{\ell=
-\infty}^{\infty}{i^\ell}J_\ell(bw,*)e^{i{\ell}s}
{=}\sum_nJ_n(w,*)(a{+}ib)^ne^{nis}.
$$

\noindent
{\bf The Generating function of Legendre polynomials}  $P_n(z)$ is 
$$
\frac{1}{\sqrt{1{-}2tz{+}t^2}}=\sum_{n=0}^{\infty}P_n(z)t^n, \quad
\text{for small }\,\,|t|.
$$
It is known that 
$P_n(z)=\frac{1}{2^nn!}\frac{d^n}{dz^n}(z^2{-}1)^{n}$.
Hence 
$\frac{1}{\sqrt{1{-}2t(z{+}a){+}t^2}}=\sum_n\frac{1}{2^nn!}\frac{d^n}{da^n}((z{+}a)^2{-}1)^{n}t^n$
is viewed as the Taylor expansion of the l.h.s. 
Using Laplace transform, we rewrite the l.h.s, and we see     
$$
\frac{1}{\sqrt{1{-}2t(z{+}a){+}t^2}}
=
\frac{1}{\sqrt{\pi}}
\int_0^{\infty}\frac{1}{\sqrt{s}}e^{-s(1{-}2t(z{+}a){+}t^2)}ds =
\sum_{n=0}^{\infty}P_n(z{+}a)t^n.
$$ 
This implies also that  
\begin{equation}\label{laplace}
\frac{d^n}{dt^n}\Big|_{t=0}\frac{1}{\sqrt{\pi}}
\int_0^{\infty}\frac{1}{\sqrt{s}}e^{-s(1{-}2t(z{+}a){+}t^2)}ds
=\frac{1}{2^n}\frac{d^n}{da^n}((z{+}a)^2{-}1)^{n}.
\end{equation}

Replacing the exponential function in the integrand by the
$*$-exponential function, we define $*$-Legrendre polynomial by 
$$
\frac{1}{\sqrt{\pi}}
\int_0^{\infty}\frac{1}{\sqrt{s}}e_*^{-s(1{-}2t(w{+}a){+}t^2)}ds =
\sum_{n=0}^{\infty}P_n(w{+}a,*)t^n.
$$
As ${:}e_*^{-s(1{-}2t(w{+}a){+}t^2)}{:}_{\tau}=e^{\tau s^2t^2}e^{-s(1{-}2t(w{+}a){+}t^2)}$,
we assume that ${\rm{Re}}\,\tau<0$ so that the integral converges.
$$
\frac{1}{\sqrt{\pi}}
\int_0^{\infty}\frac{1}{\sqrt{s}}e^{\tau s^2t^2}e^{-s(1{-}2t(w{+}a){+}t^2)}ds =
\sum_{n=0}^{\infty}P_n(w{+}a,\tau)t^n,\quad P_n(w{+}a,\tau)={:}P_n(w{+}a,*){:}_{\tau}.
$$
As the variable $z$ is used formally in \eqref{laplace}, the same
formula as in \eqref{laplace} holds for $*$-exponential functions. i.e. 
$$
\frac{d^n}{dt^n}\Big|_{t=0}\frac{1}{\sqrt{\pi}}
\int_0^{\infty}\frac{1}{\sqrt{s}}e_*^{-s(1{-}2t(w{+}a){+}t^2)}ds
=\frac{1}{2^n}\frac{d^n}{da^n}((w{+}a)_*^2{-}1)_*^{n}.
$$
By this trick we see that  
$$
\frac{1}{\sqrt{\pi}}
\int_0^{\infty}\frac{1}{\sqrt{s}}e_*^{-s(1{-}2t(w{+}a){+}t^2)}ds
=
\sum_{n=0}^{\infty}P_n(w{+}a,*)t^n
=\sum_{n=0}^{\infty}\frac{1}{2^nn!}\frac{d^n}{dz^n}((w{+}a)_*^2{-}1)_*^{n}t^n.
$$

\bigskip
Generating functions for Bernoulli numbers, Euler numbers and Laguerre
polynomials will be mentioned in later sections, for there are some 
other problems for the treatment. 
  
\subsection{Jacobi's theta functions, and  Imaginary transformations} 

For arbitrary $a{\in}{\mathbb C}$, consider the $*$-exponential function 
$e_*^{t(w+a)}$. Since 
${:}e_*^{t(w+a)}{:}_{\tau}=e^{\frac{\tau}{4}t^2}e^{t(w+a)},$
by supposing $\rm{Re}\,\tau <0$, this is rapidly decreasing on
${\mathbb R}$. Hence we see that both 
$$
\int_{-\infty}^{\infty}{:}e_*^{t(w+a)}{:}_{\tau}dt,\quad 
\sum_{n=-\infty}^{\infty}{:}e_*^{n(w+a)}{:}_{\tau}
$$
absolute converge on every compact domain in $w$ 
to give entire functions of $w$. 

In this section, we treat first a special case
$\theta(w,*)=\sum_ne_*^{2inw}$ under the condition  
$\rm{Re}\,\tau >0$.
If we set $q{=}e^{-\tau}$, the $\tau$-expression 
$\theta(w,\tau){=}{:}{\theta}(w,*){:}_{\tau}$ is given by  
$\theta(w,\tau){=}\sum_{n{\in}{\mathbb Z}}q^{n^2}e^{2niw}.$
This is Jacobi's elliptic $\theta$-function $\theta_3(w,\tau)$.

Furthermore, Jacobi's 
elliptic theta functions $\theta_i, i{=}1,2,3,4$ are 
$\tau$-expressions of bilateral geometric series of $*$-exponential functions as 
follows (cf. \cite{AAR})¡§
\begin{equation}
 \label{eq:jacobi}
\begin{aligned}
{\theta}_1(w,*)=&
\frac{1}{i}\sum_{n=-\infty}^{\infty}(-1)^ne_*^{(2n{+}1)iw},\qquad 
{\theta}_2(w,*)=  \sum_{n=-\infty}^{\infty}e_*^{(2n{+}1)iw},\\
{\theta}_3(w,*)=&
 \sum_{n=-\infty}^{\infty}e_*^{2niw},\qquad\qquad\qquad\quad
{\theta}_4(w,*)=\sum_{n=-\infty}^{\infty}(-1)^ne_*^{2niw}
\end{aligned}  
\end{equation}

\smallskip 
This fact has been mentioned first 
in \cite{om6}, and no further investigation of this fact has been done. 

The exponential law $e_*^{aw{+}s}=e_*^{aw}e^{s}$ for $s{\in}
{\mathbb C}$ gives that  
${\theta}_i(w,*)$ are $2\pi$-periodic. (Precisely,
$\theta_1(w,*)$, $\theta_2(w,*)$ are alternating $\pi$-periodic, and 
$\theta_3(w,*)$, $\theta_4(w,*)$ are $\pi$-periodic.)  
Furthermore the exponential law \eqref{eq:explaw1}
gives the trivial identities 
$$
e_*^{2iw}{*}{\theta}_i(w,*)={\theta}_i(w,*),\,\, (i{=}2,3),
\quad
e_*^{2iw}{*}{\theta}_i(w,*)=-{\theta}_i(w,*),\,\, (i{=}1,4).
$$

For every $\tau$ such that ${\rm{Re}}\,\tau >0$, 
$\tau$-expressions of these are given by using 
${:}e_*^{2iw}{:}_{\tau}=e^{-\tau}e^{2iw}$ and 
\eqref{eq:prodexp22} as follows:
\begin{equation}
\begin{aligned}
&e^{2iw-\tau}\theta_i(w{+}i\tau, \tau){=}\theta_i(w,\tau),
 \,\,(i=2,3),\\ 
&e^{2iw-\tau}\theta_i(w{+}i\tau, \tau){=}{-}\theta_i(w,\tau), \,\,(i=1,4).
\end{aligned}
\end{equation}
$\theta_i(w;*)$ is a parallel section defined on the open 
right half-plane, but 
the expression parameter $\tau$ turns out to give the 
quasi-periodicity with 
the exponential factor $e^{2iw-\tau}$. 

Noting that $(e_*^{2iw}{-}1){*}\theta_3(w, *){=}0$ in the 
computation of $*_{\tau}$-product, we have 
\begin{prop}\label{easy00}
If $f{\in}H\!ol(\mathbb C)$ satisfies $f(w{+}\pi){=}f(w)$ and
${:}(e_*^{2iw}{-}1){:}_\tau{*_\tau}f{=}0$, then 
$f{=}c{:}\theta_3(w, *){:}_\tau,\,\, c{\in}{\mathbb C}.$
\end{prop} 

\noindent
{\bf Proof}\, By the periodicity, the Fourier expansion theorem gives 
$f(w){=}\sum a_ne^{2in w}$, but by the formula of $*$-exponential functions, this is 
rewritten as $f(w)={:}\sum c_ne_*^{2in w}{:}_\tau$. 
This gives the result, for the second identity gives 
that $c_{n+1}=c_n$. \hfill $\square$

\subsubsection{Two different inverses of an element and 
$*$-delta functions}\label{sect-invss}

The convergence of bilateral geometric series for a $*$-exponential
functions give a  little strange features. Note that if 
${\rm{Re}}\,\tau>0$, then $\tau$-expressions of  
$\sum_{n=0}^{\infty}e_*^{2niw}$ and    
$-\sum_{n=-\infty}^{-1}e_*^{2niw}$ both converge in $H\!ol(\mathbb C)$ 
to give inverses of the element ${:}(1{-}e_*^{2iw}){:}_{\tau}$, and 
$\theta_3(w,\tau)$ is the difference of these inverses. 
We denote these inverses by using short notations:
$$
(1{-}e_*^{2iw})^{-1}_{*+}{=}\sum_{n=0}^{\infty}e_*^{2niw},\quad
(1{-}e_*^{2iw})^{-1}_{*-}{=}-\sum_{n=1}^{\infty}e_*^{-2niw},\quad 
(1{-}e_*^{-2iw})^{-1}_{*+}{=}\sum_{n=0}^{\infty}e_*^{-2niw}
$$
Apparently, this breaks associativity:
$$
\Big((1{-}e_*^{2iw})^{-1}_{*+}{*_\tau}(1{-}e_*^{2iw})\Big){*_\tau}(1{-}e_*^{2iw})^{-1}_{*-}
\not=
(1{-}e_*^{2iw})^{-1}_{*+}{*_\tau}\Big((1{-}e_*^{2iw}){*_\tau}(1{-}e_*^{2iw})^{-1}_{*-}\Big).
$$

\medskip
Similarly, $\theta_4(w,*)$ is the difference of two inverses of 
$1{+}e_*^{2iw}$
$$
(1{+}e_*^{2iw})^{-1}_{*+}{=}\sum_{n=0}^{\infty}(-1)^ne_*^{2niw},\quad
(1{+}e_*^{2iw})^{-1}_{*-}{=}-\sum_{n=1}^{\infty}(-1)^ne_*^{-2niw}{=}
(1{+}e_*^{-2iw})^{-1}_{*+}{-}1.
$$
Note also
$2e_*^{iw}{*}\sum_{n\geq 0}(-1)^ne_*^{2inw}$ and 
$2e_*^{-iw}{*}\sum_{n\geq 0}(-1)^ne_*^{-2inw}$ are 
both ${*}$-inverses of $\frac{1}{2}(e_*^{iw}{+}e_*^{-iw})$. We denote 
these by 
$(\cos_*w)^{-1}_{*+}, \quad  (\cos_*w)^{-1}_{*-}.$
Then, we see  
$$
2i\theta_1(w,*){=}(\cos_*w)^{-1}_{*+}-(\cos_*w)^{-1}_{*-}.
$$
Every $\theta_i(w,*)$ is written by differences of two different inverses.

Next, we note the similar phenomenon as above 
for the generator of the algebra: 

\begin{prop}\label{geninverse}
If ${\rm{Re}}\,\tau{>}0$, then for every 
$a{\in}{\mathbb C}$, the integrals 
$
i\!\int_{-\infty}^0{:}e_*^{it(a+w)}{:}_\tau dt, \,\,  
{-}i\!\int_{0}^{\infty}{:}e_*^{it(a+w)}{:}_\tau dt
$ 
converge in $H\!ol(\mathbb C)$ to give inverses of $a{+}w$. 
\end{prop}
Denote these inverses by 
$$
{:}(a{+}w)_{*+}^{-1}{:}_{\tau}\,
{=}i\int_{-\infty}^0{:}e_*^{it(a+w)}{:}_{\tau}dt,\quad  
{:}(a{+}w)_{*-}^{-1}{:}_{\tau}\,
{=}{-}\!i\int_{0}^{\infty}{:}e_*^{it(a+w)}{:}_{\tau}dt, 
\quad {\rm{Re}}\,\tau >0. 
$$
The difference of these two inverses is given by  
\begin{equation}\label{deltafunc}
(a{+}w)_{*+}^{-1}{-}(a{+}w)_{*-}^{-1}{=}
{i}\int_{-\infty}^{\infty}e_*^{it(a+w)}dt,\quad {\rm{Re}}\,\tau >0.
\end{equation}
The right hand side may be viewed as a $\delta$-function 
in the world of $*$-functions. 
Set   
\begin{equation}\label{stardelta}
\delta_*(a{+}w)=\frac{1}{2\pi}\int_{-\infty}^{\infty}e_*^{it(a+w)}dt\quad {\rm{Re}}\,\tau >0
\end{equation}
and we call \eqref{stardelta} the $*$-$\delta$ function. We see easily that $(a{+}w){*}\delta_*(a{+}w)=0$.  
Note that  $(a{+}w)_{*+}^{-1}{+}ci\delta_*(a{+}w)$ 
gives the inverse of $a{+}w$ for any constant $c$.

In the ordinary calculus, 
$\int_{-\infty}^{\infty}e^{it(a+x)}dt{=}2\pi\delta(a{+}x)$ is 
not a function but a distribution. On the contrary,
in the world of $*$-functions, 
the $\tau$-expression ${:}\delta_*(a{+}w){:}_\tau$  
of $\delta_*(a{+}w)$ is an entire function: 
\begin{equation}\label{eq:Fouriertrsf}
{:}\delta_*(a{+}w){:}_\tau=
\frac{1}{2\pi}\int_{-\infty}^{\infty}e^{-\frac{1}{4}t^2\tau}e^{it(a{+}w)}dt{=}
\frac{1}{\sqrt{\pi\tau}}e^{-\frac{1}{\tau}(a{+}w)^2}, \quad 
{\rm{Re}}\,\tau >0, \quad a{\in}{\mathbb C}
\end{equation}

\subsubsection{Jacobi's imaginary transformations}\label{thetaRiem}
By the formula \eqref{eq:Fouriertrsf}, we see  
the following series 
$$
\begin{aligned}
&\tilde\theta_1(w ,*){=}
\sum_n(-1)^n\delta_*(w {+}\frac{\pi}{2}{+}\pi n), \quad 
\tilde\theta_2(w ,*){=}\sum_n(-1)^n\delta_*(w{+}\pi n)\\
&\tilde\theta_3(w ,*){=}\sum_n \delta_*(w {+}\pi n), 
\qquad\qquad\quad
\tilde\theta_4(w ,*){=} 
\sum_n \delta_*(w{+}\frac{\pi}{2}{+}\pi n),
\end{aligned}
$$
converge in the $\tau$-expression for ${\rm{Re}}\,\tau>0$. 
These may be viewed as $\pi$-periodic/$\pi$-alternating periodic 
${*}$-delta function on ${\mathbb R}$. 
As $e^{2\pi in}{=}1$, we have identities 
$$
e_*^{2iw }{*}\tilde\theta_i(w ,*){=}\tilde\theta_i(w ,*), \quad 
(i=2,3),\quad 
e_*^{2iw }{*}\tilde\theta_i(w ,*){=}{-}\tilde\theta_i(w ,*),\quad 
(i=1,4). 
$$
By a slight modification of Proposition\,\ref{easy00}, we have 
$\theta_i(w ,*)=\alpha_i\tilde\theta_i(w ,*), \,\,\alpha_i\in{\mathbb C}.$
Note that $\alpha_i$ does not depend on the expression parameter $\tau$. 
Taking the $\tau$-expressions of both sides at $\tau{=}\pi$ and  
setting $w{=}0$, we have $\alpha_i=1/2$.

\begin{prop}
$\theta_i(w ,*)=\frac{1}{2}\tilde\theta_i(w ,*)$ for 
$i=1\sim 4$. 

The Jacobi's imaginary transformation is given by 
taking the $\tau$-expression of these identities.
\end{prop}

This may be proved directly by the following manner: Since  
$f(t){=}\sum_ne_*^{2(n+t)iw }$ is periodic function of period $1$, 
Fourier expansion formula gives 
$$
f(t){=}\sum_m \int_0^{1}f(s)e^{-2\pi ims}ds e^{2\pi imt}, \quad 
\theta_3(w,*){=}f(0){=}
\sum_m \int_0^{1}(\sum_ne_*^{2(n+s)iw })e^{-2\pi ims}ds.
$$  
Since $e^{-2\pi ims}=e^{-2\pi im(s+n))}$, we have 
\begin{equation*}\label{thetarel}
f(0){=}\sum_m \int_0^{1}
(\sum_ne_*^{2(n+s)iw }e^{-2(n+s)i\pi m})ds 
=
\sum_m \int_{-\infty}^{\infty}e_*^{2si(w +\pi m)}ds
{=}
\frac{1}{2}\sum_m \delta_*(w +\pi m).
\end{equation*}

\medskip 
Hence \eqref{eq:Fouriertrsf} gives   
\begin{equation}\label{thetarel}
\begin{aligned}
&\theta_3(w,{\tau})
{=}\frac{2\pi}{2}{:}\sum_n \delta_*(w{+}\pi n){:}_{\tau}
= \sqrt{\frac{\pi}{\tau}}
  \sum_n e^{-\frac{1}{\tau}(w{+}\pi n)^2}\\
&{=}
\sqrt{\frac{\pi}{\tau}}
e^{-\frac{1}{\tau}w ^2}
\sum_n e^{-\pi^2n^2\tau^{-1}-2\pi n\tau^{-1}w}
=
\sqrt{\frac{\pi}{\tau}}
e^{-\frac{1}{\tau}w ^2}
\theta_3(\frac{\pi w}{i\tau},\frac{\pi^2}{\tau})
\end{aligned}
\end{equation}
This is remarkable since a relation between two different 
expressions (viewpoints) are explicitly given. 

In particular, Jacobi's theta relation is 
obtained by setting $w=0$ in \eqref{thetarel}:
\begin{equation}\label{Jacobireleq}
\theta_3(0,{\tau})=
\sqrt{\frac{\pi}{\tau}}\theta_3(0,\frac{\pi^2}{\tau}).
\end{equation}
This will be used to obtain the functional identities of the $*$-zeta function 
in a forthcomming paper. 
%
%
\subsection{Calculus of inverses}

We first note that the method of constant variation creates many inverses 
of a single element. By the product formula  
$(a{+}w)*$, $a{\in}{\mathbb C}$, is viewed as a linear operator of 
$H\!ol({\mathbb C})$ into itself. If $\tau\not=0$,
$(a{+}w)*_{\tau}f(w)=0$ 
gives a differential equation 
$(a{+}w)f(w){+}\frac{\tau}{2}\partial_{w}f(w)=0.$
Solving this, we have  
$(a{+}w)*_{\tau}Ce^{-\frac{1}{\tau}(a{+}w)^2}= 
Ce^{-\frac{1}{\tau}(a{+}w)^2}{*_\tau}(a{+}w)=0.$ 
The method of constant variation gives a function 
$g_a(w)$ such that 
$(a{+}w){*_{\tau}}g_a(w)=g_a(w){*_\tau}(a{+}w)=1.$  
Thus, we have   
\begin{equation}
g_{a}(w)=
\frac{2}{\tau}
\int_0^{1}e^{\frac{1}{\tau}((a{+}wt)^2{-}(a{+}w)^2)}wdt{+}C
e^{-\frac{1}{\tau}(a{+}w)^2},\quad \tau\not=0.
\end{equation}
Hence this breaks associativity 
$(e^{-\frac{1}{\tau}(a{+}w)^2}{*_\tau}(a{+}w))*_{\tau}g_a(w)\not=
e^{-\frac{1}{\tau}(a{+}w)^2}{*_\tau}((a{+}w)*_{\tau}g_a(w)).$
If $b{+}w$ has also two different $*$-inverses, then by providing
$a\not=b$, 4 elements with independent $\pm$-sign 
$$
\frac{1}{b{-}a}(\big((a{+}w)_{*\pm}^{-1}-(b{+}w)_{*\pm}^{-1}\big))
$$
give respectively ${*}$-inverses of $(a{+}w){*}(b{+}w)$.
Thus, we define ${*}$-inverse with independent $\pm$-sign by 
\begin{equation}\label{defprodinv}
(a{+}w)_{*\pm}^{-1}{*}(b{+}w)_{\pm}^{-1}=
\frac{1}{b{-}a}\big((a{+}w)_{*\pm}^{-1}-(b{+}w)_{*\pm}^{-1}\big).
\end{equation}
Then the direct computation ${*}$-product shows  for any  
$a,b\in {\mathbb C}, \,\,a\not=b$ that
$$
\Big((a{+}w)^{-1}_{*+}{-}(a{+}w)_{*-}\Big){*}
\Big((b{+}w)^{-1}_{*+}{-}(b{+}w)^{-1}_{*-}\Big)=0.
$$

\subsubsection{Half-series algebra}

It is well known that 
if a formal power series satisfies $\sum_{n=0}^{\infty}a_nz^n{=}0$, then
$a_n=0$. This is proved by setting $z{=}0$ to get $a_0=0$, and then
taking $\partial_z|_{z=0}$ to get $a_1{=}0$ and so on.  
Hence this method cannot be applied to formal power series 
$\sum_{n=0}^{\infty}a_ne_*^{niw}$. 

We suppose ${\rm{Re}}\,\tau>0$ throughout this subsection. 
A formal power series $z^{\ell}\sum_{n=0}^{\infty}a_nz^n$, 
$\ell{\in}\mathbb Z$, is called a {\it convergent power series}, if 
$\sum_{n=0}^{\infty}a_nz^n$ has a positive radius of convergence.
Proposition\,\ref{entirelevel} shows that if 
$z^{\ell}\sum_{n=0}^{\infty}a_nz^n$ is a convergent power series, then 
$f(w)={:}e_*^{\ell iw}{*}\sum_{n=0}^{\infty}a_ne_*^{niw}{:}_{\tau}$ is an 
entire function of $w$. Hence if $f(w)=0$, then Proposition\,\ref{ketugoprodexp22} gives 
$\sum_{n=0}^{\infty}a_ne_*^{niw}{:}_{\tau}=0$, and $a_0=0$ by taking
$w\to i\infty$. Thus the repeated use of Proposition\,\ref{ketugoprodexp22} gives 
all $a_n=0$.

Note that the product of two convergent power series is a convergent
power series. 
If $z^{\ell}\sum_{n=0}^{\infty}a_nz^n$,($\ell \in
\mathbb Z$) is a convergent power series, then its inverse 
$(z^{\ell}\sum_{n=0}^{\infty}a_nz^n)^{-1}$ obtained by 
the method of indeterminate constants is also a convergent power series. 
We denote by 
${\mathfrak H}_+$ be the space of power series 
${:}e_*^{\ell iw}{*}\sum_{n=0}^{\infty}a_ne_*^{niw}{:}_{\tau}$
made by convergent power seris  $z^{\ell}\sum_{n=0}^{\infty}a_nz^n$.
We call ${\mathfrak H}_+$ the half-series algebra. Its fundamental
property is 
\begin{thm}\label{nicefield}
$({\mathfrak H}_+, {*_{\tau}})$ is a topological field of $2\pi$
periodic entire functions of $w$.  
\end{thm}

\noindent
{\bf Proof}\, is completed by showing the uniquness of the
inverse. It is reduced to show that 
$$
\sum_{n=0}^{\infty}a_ne_*^{niw}{:}_{\tau}{*_{\tau}}\sum_{k=0}^{\infty}b_ke_*^{kiw}{:}_{\tau}=0
$$
and $a_0\not=0$ gives $\sum_{k=0}^{\infty}b_ke_*^{kiw}{:}_{\tau}=0$. 
The repeated use of Proposition\,\ref{ketugoprodexp22} gives 
all $b_n=0$. \hfill $\Box$

\bigskip
\noindent
{\bf Euler numbers}\,
Recall the the generating function of Euler numbers 
$$
\frac{2}{e^{z}{+}e^{-z}}=
\frac{e^z}{1{+}e^{2z}}{+}\frac{e^{-z}}{1{+}e^{-2z}}{=}
\sum_{n=0}^{\infty}E_{2n}\frac{1}{(2n)!}z^{2n},\quad |z|<\pi.
$$ 
The l.h.s. is a convergent power series obtained by the method of indeterminate constants.
Hence by Proposition\,\ref{entirelevel} gives 
\begin{equation}\label{Euler}
e_*^{e_*^{iw}}{*}\Big(1{+}\sum_{k=0}^{\infty}{2^ke_*^{kiw}}\frac{1}{k!}\Big)^{-1}
{+}e_*^{-e_*^{iw}}{*}\Big(1{+}\sum_{k=0}^{\infty}{(-2)^ke_*^{kiw}}\frac{1}{k!}\Big)^{-1}=
\sum_{n=0}^{\infty}E_{2n}\frac{1}{(2n)!}e_*^{2niw},
\end{equation}
where $e_*^{\pm e_*^{iw}}=\sum_{\ell=0}^{\infty}\frac{(\pm 1)^{\ell}}{\ell !}
e_*^{\ell iw}$.

On the other hand, by using the formal power series of $(iw)_*^n$, 
we can compute the inverces 
$\Big(1{+}\sum_{k=0}^{\infty}\frac{(2iw)_*^k}{k!}\Big)^{-1}$, 
$\Big(1{+}\sum_{k=0}^{\infty}\frac{(-2iw)_*^k}{k!}\Big)^{-1}$  
by the method of indeterminate constants. Hence we have also 
\begin{equation}\label{Euler22}
e_*^{iw}{*}\Big(1{+}\sum_{k=0}^{\infty}{(2iw)_*^k}\frac{1}{k!}\Big)^{-1}
{+}e_*^{-iw}{*}\Big(1{+}\sum_{k=0}^{\infty}(-2iw)_*^k\frac{1}{k!}\Big)^{-1}=
\sum_{n=0}^{\infty}E_{2n}\frac{1}{(2n)!}(iw)_*^{2n}.
\end{equation} 
It is clear that the replacement $(iw)_*^k$ by $e_*^{kiw}$ gives \eqref{Euler}. 
It is very interesting to compare the l.h.s with  
$e_*^{iw}(1{+}e_*^{2iw})_{*+}^{-1}{+}e_*^{-iw}{*}(1{+}e_*^{-2iw})_{*+}^{-1}$. It
is natural to have the following 

\bigskip
\noindent
{\bf Conjecture}\,\,\,By using another expression parameter $\tau'$ such
that ${\rm{Re}}\,\tau'>0$ and ${\rm{Re}}(\tau{-}\tau')>0$, the
$\tau'$-expression of   
$e_*^{iw}(1{+}e_*^{2iw})_{*+}^{-1}{+}e_*^{-iw}{*}(1{+}e_*^{-2iw})_{*+}^{-1}$
is an entire function of $w$. Denote this by 
$$
{:}e_*^{iw}(1{+}e_*^{2iw})_{*+}^{-1}{+}e_*^{-iw}{*}(1{+}e_*^{-2iw})_{*+}^{-1}{:}_{\tau'}
=\sum_{n=0}^{\infty}a_{2n}(\tau,\tau'){:}(iw)_*^{2n}{:}_{\tau'}
$$
and regard the r.h.s as a $\tau'$-expression of the $*$-function 
$\sum_na_{2n}(\tau,\tau')(iw)_*^{2n}$. Then, the replacement
$(iw)_*^{2n}$ by $e_*^{2inw}$ gives 
$$
{:}\sum_na_n(\tau,\tau')e_*^{niw}{:}_{\tau'}=
{:}\sum_{n=0}^{\infty}E_{2n}\frac{1}{(2n)!}e_*^{2niw}{:}_{\tau{-}\tau'}.
$$

\bigskip
\noindent
{\bf Bernoulli numbers}\,\,
Recall here the generating function of Bernoulli numbers:
$$
z\Big(\frac{1}{2}{+}\frac{1}{e^z{-}1}\Big){=}
\frac{z}{2}\Big(\frac{1}{e^{z}{-}1}{-}\frac{1}{e^{-z}{-}1}\Big)
{=}\sum_{n=0}^{\infty}B_{2n}\frac{1}{(2n)!}z^{2n}. 
$$
Since $\frac{z}{e^z{-}1}$ and $\frac{-z}{e^{-z}{-}1}$ are 
computed by the method of indeterminate constants as 
$$
(\sum_n\frac{z^n}{(n+1)!})^{-1}
=\sum B_{2n}\frac{1}{(2n)!}z^{2n}{-}\frac{1}{2}z,\quad 
(\sum_n\frac{(-z)^n}{(n+1)!})^{-1}
=\sum B_{2n}\frac{1}{(2n)!}z^{2n}{+}\frac{1}{2}z,
$$
As in \eqref{Euler}, the r.h.s is a convergent power series.
Hence we have 
\begin{equation}\label{Bern}
\frac{1}{2}\Big(\sum_n\frac{e_*^{niw}}{(n+1)!}\Big)^{-1}{+}
\frac{1}{2}\Big(\sum_n\frac{-e_*^{niw}}{(n+1)!}\Big)^{-1}
=\sum_{n=0}^{\infty}B_{2n}\frac{1}{(2n)!}e_*^{2niw}
\end{equation}

\medskip
On the other hand, we have for every $\tau'$ a formal power series 
$$
{:}\frac{1}{2}\Big(\sum_n\frac{(iw)_*^n}{(n+1)!}\Big)^{-1}
{+}\frac{1}{2}\Big(\sum_n\frac{(-iw)_*^n}{(n+1)!}\Big)^{-1}{:}_{\tau'}=
\sum_{k=0}^{\infty}
B_{2k}{:}\frac{(iw)_*^{2k}}{(2k)!}{:}_{{\tau}'}
$$
where both sides are computed as formal power series of $(iw)$.  
It is clear that the replacement 
$(iw)_*^{2k}$ by $e_*^{2kiw}$ in the r.h.s gives 
$\sum_{k=0}^{\infty}B_{2k}\frac{(e_*^{2kiw}}{(2k)!}$. 
Hence, we have the same conjecture for 
$$
{:}\frac{1}{2}iw{*}\Big((e_*^{iw}{-}1)_{*+}^{-1}{-}(e_*^{-iw}{-}1)_{*+}^{-1}\Big){:}_{\tau'}
$$

\section{Srar-functions made by tempered distributions} 

Throughout this section, we assume ${\rm{Re}}\,\tau>0$. Note that 
${:}\delta_*(x{-}w){:}_{\tau}=
\frac{1}{\sqrt{\pi\tau}}e^{{-}\frac{1}{\tau}(x-w)^2}$ is rapidly decreasing.
Suppose $f(x)$ is $e^{|x|^{\alpha}}$-growth on ${\mathbb R}$ with 
$0{<}\alpha{<}2$.  
Then the integral $\int f(x){:}\delta_*(x{-}w){:}_{\tau}dx$ 
is well-defined to give an entire function w.r.t.$w$. 

The next theorem is a main tool to extend the class of 
$*$-functions via Fourier transform:
\begin{thm}
  \label{funddefm}
For every tempered distribution $f(x)$, 
the $\tau$-expression of  
$\int_{-\infty}^{\infty}f(x)\delta_*(x{-}w)dx$ is an entire function 
of $w$ whenever ${\rm{Re}}\,\tau{>}0$. In particular we see 
$\delta_*(a{-}w)=\int_{-\infty}^{\infty}\delta(x{-}a)\delta_*(x{-}w)dx.$
\end{thm} 
Although the product 
$\delta_*(x{-}w){*}\delta_*(x{-}w)$ diverges, 
the next one is important 
\begin{equation}\label{sekidlt}
\delta_*(x{-}w){*}\delta_*(x'{-}w)=
\delta(x-x')\delta_*(x'{-}w) 
\end{equation} 
in the sense of distribution. This is proved directly as follows: 
$$
\begin{aligned}
\delta_*(x{-}w)&{*}\delta_*(x'{-}w)=
(\frac{1}{2\pi})^2\iint 
e_*^{it(x{-}w)}{*}e_*^{is(x{-}w)}dtds\\
=&
(\frac{1}{2\pi})^2\iint e^{itx{+}isx'}e_*^{-i(t{+}s)w}dtds=
(\frac{1}{2\pi})^2\iint e^{is(x'{-x)}}e_*^{i\sigma(x{-}w)}dsd\sigma
=\delta(x'{-}x)\delta_*(x{-}w).
\end{aligned}
$$
For every tempered distribution $f(x)$, 
we define a $*$-function $f_*(w)$ by 
\begin{equation}\label{startemparete}
f_*(w){=}\int_{-\infty}^{\infty}f(w)\delta_*(x{-}w)dx= 
\frac{1}{\sqrt{2\pi}} 
\int_{-\infty}^{\infty}\check f(t)e_*^{-itw}dt.
\end{equation}
where $\check f(t)$ is the inverse Fourier 
transform of $f(x)$. 
As $f(x)$ is a tempered distribution, one may write  
$$
\int f(x){:}\delta_*(x{-}w){:}_{\tau}dx 
= 
\frac{1}{2\pi}
\iint f(x)e^{itx}{:}e_*^{-itw}{:}_{\tau}dtdx
$$
under the existence of a rapidly decreasing function 
${:}e_*^{-itw}{:}_{\tau}$ in the integrand. 
By the definition of Fourier transform of tempered distribution, 
one may exchange the order of integrations.  
Letting ${\check f}(t)$ be the inverse Fourier transform of 
$f(x)$, we have 
\begin{equation}\label{nicerep}
{:}\int_{\mathbb R} f(x)\delta_*(x{-}w)dx{:}_{\tau} 
= 
\frac{1}{\sqrt{2\pi}}
\int_{\mathbb R}{\check f}(t){:}e_*^{-itw}{:}_{\tau}dt
{=}{:}f_*(w){:}_{\tau}. 
\end{equation}
 
If another $*$-function is given by 
$g_*(w)=\int g(x)\delta_*(x{-}w)dx$, we define the 
product by 
\begin{equation}\label{productstar}
f_*(w){*}g_*(w)=
\int_{-\infty}^{\infty}f(w)g(w)\delta_*(x{-}w)dx
=\frac{1}{\sqrt{2\pi}}\int
\big(\frac{1}{\sqrt{2\pi}}\int {\check f}(t{-}\sigma)
{\check g}(\sigma)d\sigma\big)e_{*}^{-itw}dt,
\end{equation} 
if  $f(x)g(x)$ is defined as a tempered distribution or the convolution product 
${\check f}\bullet{\check g}(t){=}
\frac{1}{\sqrt{2\pi}}\int {\check f}(t{-}\sigma){\check g}(\sigma)d\sigma$ 
is defined as a tempered distribution. 
Hence \eqref{productstar} may be viewed as an integral representation of
the intertwiner $I_0^{\tau}(f(x))=f_*(w)$.
If $f(x)$ is a slowly increasing function (a function with the value
at each point $x\in{\mathbb R}$ and a tempered distributiuon),  
applying \eqref{productstar} to the case $g_*(w)=\delta_*(a{-}w)$ gives 
\begin{equation}\label{funnystardelta}
f_*(w)*\delta_*(a{-}w)=
\int f(x)\delta(a{-}x)\delta_*(x{-}w)dx
{=}f(a)\delta_*(a{-}w).
\end{equation}

\subsection{Several applications}
Note that $\frac{1}{a-w}$, $a\not\in{\mathbb R}$, is a slowly
increasing function. It is not hard to verify 
$$
\int\frac{1}{a-x}\delta_*(x{-}w)dx=
\left\{
\begin{matrix}
\medskip
(a{-}w)_{*+}^{-1}& {\rm{Im}\,a}<0\\
(a{-}w)_{*-}^{-1}& {\rm{Im}\,a}>0
\end{matrix}
\right.,\quad {\rm{Re}}\,\tau>0.
$$

For ${\rm{Re}}\,\tau>0$, we define  
$$
Y_*(w)=\lim_{\e\downarrow 0}\int_{\e}^{\infty}\delta_*(x{-}w)dx, \quad 
Y_*(-w)=\lim_{\e\downarrow 0}\int_{-\infty}^{-\e}\delta_*(x{-}w)dx.
$$
It is clear that 
$$
\partial_w\lim_{\e\downarrow 0}\int_{\e}^{\infty}\delta_*(x{-}w)dx=
-\lim_{\e\downarrow 0}\int_{\e}^{\infty}\partial_x\delta_*(x{-}w)dx=
\delta_*({-}w)=\delta_*(w).
$$
Using \eqref{sekidlt} we have 
$Y_*(w){*}Y_*(w)=Y_*(w), \quad Y_*(w){*}Y_*(-w)=0,\quad 
Y_*(w){+}Y_*(-w)=\int_{\mathbb R}\delta_*(x{-}w)dx=1.$

We define 
$$
{\rm{sgn}}_*(w)=Y_*(w){-}Y_*(-w).
$$
It is easy to see that  
${\rm{sgn}}_*(w){*}{\rm{sgn}}_*(w)=Y_*(w){+}Y_*(-w)=1, \quad {\rm{sgn}}_*(w){+}{\rm{sgn}}_*(-w)=0. $

\bigskip
Since $\delta_*(z{-}w)$ is holomorphic in $z$, Cauchy 
integral theorem gives that every contour integral vanishes, but we
see easily  for every simple closed curve $C$
$$
\frac{1}{2\pi i}\int_C \frac{1}{z}\delta_*(z{-}w)dz=\delta_*(w),\quad
{\rm{Re}}\tau >0.
$$  
\medskip
Note that ${\rm{v.p.}}\frac{1}{x}$, $\rm{Pf}.x^{-m}, m\in \mathbb N$ are tempered
distribution which are not functions, but their Fourier transform may
be viewed as slowly increasing functions. Hence we see 
$$
\begin{aligned}
{\rm{v.p.}}\int_{\mathbb R}\frac{1}{x}\delta_*(x{-}w)dx&= 
\frac{-i}{\sqrt{2\pi}}
\int_{\mathbb R}\!\sqrt{\frac{\pi}{2}}\,{\rm{sgn}}(t)e_*^{-itw}dt
=\frac{1}{2}(w_{*+}^{-1}{+}w_{*-}^{-1})\\ 
{\rm{Pf}.}\int_{\mathbb R}x^{-m}\delta_*(x{-}w)dx&=
\frac{-i}{2}
\int_{\mathbb R}\frac{1}{(m{-}1)!}(-it)^{m{-}1}{\rm{sgn}}(t)e_*^{-itw}dt
=(-1)^{m-1}\frac{1}{2}(w_{*+}^{-m}{+}w_{*-}^{-m}).
\end{aligned}
$$

\subsubsection{Periodical distributions}
A tempered distribution $f(x)$ is called a $2\pi$-periodic 
tempered distribution, if $f(x)$ satisfies $f(x{+}2\pi)=f(x)$.  
For every distribution $f(x)$ with compact support, the infinite sum 
$\sum_nf(x{+}2\pi n)$ is a $2\pi$-periodic tempered distribution. 
The fundamental relation between  
$2\pi$-periodic tempered distributions and 
Fourier series is 
\begin{equation}\label{periodic}
\sum_n\delta_*(a{+}2\pi n{+}w)=\sum_ne_*^{in(a{+}w)}.
\end{equation}
A continuous function $f(x)$ on $[-\pi,\pi]$ extends to a (not continuous) 
$2\pi$-periodic function ${\tilde f}_{\pi}(x)$ to give a 
$2\pi$-periodic tempered distribution, where 
$$
{\tilde f}_{\pi}(x)=\frac{1}{2\pi}\sum_n(\int_{-\pi}^{\pi}f(s)e^{-ins}ds)e^{inx}
=\sum_na_ne^{inx}
$$
Hence 
\begin{equation}\label{perperper}
{\tilde f}_{\pi *}(w)=\int_{\mathbb R}{\tilde f}_{\pi}(x)\delta_*(x{-}w)dx=\sum a_ne_*^{inw}.
\end{equation}

\section{Star-exponential function of $w_*^2$}\label{expw2}
As we have seen, the ${*}$-exponential function $e_*^{sh_*(w)}$ is 
very naive for the order of $h(x)$ is less than $2$. 
In this section, we treat the $*$-exponential function of quadratic form $w_*^2$. 
As $e^{-tx^2}$ is a slowly increase function of $x$ for
${\rm{Re}}\,t\geq 0$, the integral $\int_{\mathbb R}e^{-tx^2}\delta_*(x{-}w)dx$
defines a semigroup $e_*^{-tw_*^2}$ under the expression parameter ${\rm{Re}}\tau>0$. 
Noting that 
${:}{w}_*^2{:}_{\tau}={w}^2{+}\frac{\tau}{2}$ in the 
$\tau$-expression, we now define 
the star-exponential function of ${w}_*^2$ 
by the real analytic solution of the evolution equation 
\begin{equation}\label{expdiffeq}
\frac{d}{dt}f_t={:}{w}_{*}^2{:}_{\tau}{*_{\tau}}f_t, 
\quad  f_0 = 1.
\end{equation} 
That is in precise
$\frac{d}{dt}f_t=
\frac{\tau^2}{4}f''_t{+}{\tau}wf'_t
{+}({w}^2{+}\frac{\tau}{2})f_t,\,\, f_0=1.$
To solve this, we set 
${:}f_t{:}_{\tau}=g(t)e^{h(t){w}^2}$
by taking the uniqueness of real analytic solution in mind. 
Then, we have a system of ordinary differential equations: 
$$
\left\{
\begin{aligned}
&\frac{d}{dt}h(t)=(1{+}\tau h(t))^2, \qquad\qquad h(0)=0\\
&\frac{d}{dt}g(t)=
\frac{1}{2}(\tau^2h(t){+}\tau)g(t),\qquad g(0)=1.  
\end{aligned}
\right. 
$$
The solution 
${:}e_*^{{t}{w}_*^2}{:}_{\tau}$ is given by   
\begin{equation}
 \label{eq:expquad}
{:}e_*^{{t}{w}_*^2}{:}_{\tau}=
\frac{1}{\sqrt{1{-}\tau t}}\,
e^{\frac{t}{1{-}\tau t}{w}^2}, 
\text{ for }\forall \tau, \,\, t\tau\not=1,\quad
{\text{(double valued)}}.
\end{equation}
It is rather surprising that the solution has 
a branching singular point, and hence this does not form a 
complex one parameter group whenever $\tau\not=0$ is fixed.  
Moreover, the solution is double valued w.r.t. the variable $t$. 
This solution is obtained also via the intertwiner 
$I_0^{\tau}e^{tw^2}$ (cf.\eqref{2-to-2}). 
Note here that there is no restriction for $\tau$. 
$e_*^{tw_*^2}$ is obtained for every $\tau$

\medskip
\noindent
{\bf Generating function  of Laguerre polynomials}\,\,
$L_n^{(\alpha)}(x)$ is given as follows: 
$$
\frac{1}{(1{-}t)^{\alpha+1}}e^{-\frac{t}{1{-}t}x}= 
\sum_{n\geq 0}L_n^{(\alpha)}(x)t^n, 
\quad (|t|<1).
$$

If $\alpha{=}-\frac{1}{2}$, 
this is the $\tau=-1$ expression of $e_*^{-tw_*^2}$, i.e.
$$
{:}e_*^{-tw_*^2}{:}_{{-}1}
{=}\frac{1}{(1{-}t)^{\frac{1}{2}}}e^{{-}\frac{t}{1{-}t}w^2}
{=}\sum_{n\geq 0}L_n^{(-\frac{1}{2})}(w^2)t^n.
$$
Keeping these in mind, we define $*$-Laguerre polynomials 
$L_n(w^2,\tau)={:}L_n(w^2,*){:}_{\tau}$ by 
\begin{equation}\label{Laglag}
e_*^{tw_*^2}=
\sum_n L_n^{(-\frac{1}{2})}(w^2,*)\frac{1}{n!}t^n,\quad 
L_n^{(-\frac{1}{2})}(w^2,\tau)=
\frac{d^n}{dt^n}\Big|_{t=0}
\frac{1}{(1{-}t\tau)^{\frac{1}{2}}}
e^{\frac{1}{\tau(1{-}t\tau)}w^2}e^{-\frac{1}{\tau}w^2}. 
\end{equation}
As $t=0$ is a regular point, these are welldefined, and the
exponential law gives 
$$
L_n^{(-\frac{1}{2})}(w^2,*)=\sum_{k{+}\ell=n}L_k^{(-\frac{1}{2})}(w^2,*)
{*}L_\ell^{(-\frac{1}{2})}(w^2,*).
$$
Note that setting $x=w^2$, 
$\frac{d}{dt}\frac{x^{\alpha{-}1}}{(1{-}t\tau)^{\alpha}}e^{\frac{1}{\tau(1-t\tau)}}
=
\frac{d}{dx}\frac{1}{\tau}
\frac{x^{\alpha}}{(1{-}t\tau)^{\alpha{+}1}}e^{\frac{1}{\tau(1-t\tau)}}.$
Using this, we see that 
$$
\frac{d^n}{dt^n}\Big|_{t=0}
\frac{1}{(1{-}t\tau)^{\frac{1}{2}}}
e^{\frac{1}{\tau(1{-}t\tau)}w^2}e^{-\frac{1}{\tau}w^2}
=\Big(\tau^{-n}\frac{d^n}{dx^n}(x^{\frac{1}{2}{+}n}e^{\frac{1}{\tau}x})\Big)
x^{-\frac{1}{2}}e^{-\frac{1}{\tau}x}.
$$
It follows that setting $x=w^2$
$$
L_n^{(-\frac{1}{2})}(w^2,\tau)=
\frac{1}{n!}\Big(\tau^{-n}\frac{d^n}{dx^n}(x^{\frac{1}{2}{+}n}e^{\frac{1}{\tau}x}\Big)
x^{-\frac{1}{2}}e^{-\frac{1}{\tau}x}
$$

As in the case of Hermite polynomials, this formula is used to 
to obtain the orthogonality of $\{L_n^{(-\frac{1}{2})}(w^2,\tau)\}_n$
restricted $x=w^2$ to the real axis and supposing ${\rm{Re}}\tau<0$.  
Namely, we want to show  
$$
\int_{\mathbb R}x^{\frac{1}{2}}e^{\frac{1}{\tau}x}
L_n^{(-\frac{1}{2})}(x,\tau)L_m^{(-\frac{1}{2})}(x,\tau)dx=\delta_{n,m}.
$$
First note that $L_n(x,\tau)$ is a polynomial of degree $n$, and
$$
\int_{\mathbb R}x^{\frac{1}{2}}e^{\frac{1}{\tau}x}L_n^{(-\frac{1}{2})}(x,\tau)L_m^{(-\frac{1}{2})}(x,\tau)dx=
\int_{\mathbb R}\frac{1}{\tau^n}\frac{1}{n!}\Big(\frac{d^n}{dx^n}
x^{\frac{1}{2}{+}n}e^{\frac{1}{\tau}{x}}\Big)L_m^{(-\frac{1}{2})}(x,\tau)dx.
$$
If $n\not= m$, one may suppose $n>m$. 
Hence this vanishes by the integration by parts $n$ times.

For the case $n=m$, recalling $L_n^{(-\frac{1}{2})}(x,\tau)$ is a
polynomial of degree $n$, and 
taking $\frac{d^n}{dx^n}$ of both sides of the second equality
of \eqref{Laglag}, we have 
$$
\frac{d^n}{dx^n}L_n^{(-\frac{1}{2})}(x,\tau)
=\frac{1}{n!}\frac{d^n}{dt^n}\Big|_{t=0}\frac{d^n}{dx^n}
\frac{1}{(1{-}t\tau)^{\frac{1}{2}}}e^{\frac{t}{1{-}t\tau}x}
=\frac{1}{n!}\frac{d^n}{dt^n}\Big|_{t=0}
\frac{t^n}{(1{-}t\tau)^{\frac{1}{2}{+}n}}e^{\frac{t}{1{-}t\tau}x}.
$$
But the last term does not contain $x$ for this must be degree $0$. Hence 
$$
\frac{d^n}{dx^n}L_n^{(-\frac{1}{2})}(x,\tau)
=\frac{1}{n!}\frac{d^n}{dt^n}\Big|_{t=0}
\frac{t^n}{(1{-}t\tau)^{\frac{1}{2}{+}n}}=1.
$$

\medskip  
In spite of double valued nature of $e_*^{t{w}_*^2}$, if a continuous curve $C$ does not hit
singular points, then ${:}e_*^{tw_*^2}{:}_{\tau}$ can be treated as a 
continuous function on $C$. For instance, one can treat the integral 
$\int_C{:}e_*^{tw_*^2}{:}_{\tau}dt$ without ambiguity.  
The uniqueness of real analytic solution gives the exponential law 
$e_{*}^{sw_*^2}{*}e_{*}^{tw_*^2}=e_{*}^{(s+t)w_*^2}$:
$$
\frac{1}{\sqrt{1{-}\tau s}}\,e^{\frac{s}{1{-}\tau s}{w}^2}
{*_{\tau}}
\frac{1}{\sqrt{1{-}\tau t}}\,e^{\frac{t}{1{-}\tau t}{w}^2}
=
\frac{1}{\sqrt{1{-}\tau(s{+}t)}}\,e^{\frac{s{+}t}{1{-}\tau(s{+}t)}{w}^2}.
$$ 
Indeed this holds through calculations such as 
$\sqrt{a}\sqrt{b}=\sqrt{ab}$, $\sqrt{a}/\sqrt{a}=\sqrt{1}=\pm 1$. 

Similarly, we have the exponential law  $e^{s}{*}e_{*}^{tw_*^2}=e_{*}^{s+tw_*^2}$
with an ordinary scalor exponential function $e^{s}$.

\subsection{Intertwiners are 2-to-2 mappings}
 Recall that the intertwiner $I_{\tau}^{\tau'}$ is defined by 
$e^{\frac{1}{4}(\tau'{-}\tau)\partial^2_w}$. 
For the case of exponential functions of quadratic forms,
this is treated by solving the evolution equation 
$\frac{d}{dt}f_t(w)=\partial_w^2 f(w),\quad f_0(w)=ce^{aw^2}$.
Setting $f_t=g(t)e^{q(t)w^2}$, this equation is changed into 
$$
\left\{
\begin{matrix}
\frac{d}{dt}q(t)= 4q(t)^2  & q(0)=a \\
{}                         &{}\\
\frac{d}{dt}g(t)= 2g(t)q(t)& g(0)=c \\
\end{matrix}
\right.
$$
Solving this to get 
$g(t)e^{q(t)w^2}=
\frac{c}{\sqrt{1-4ta}}e^{\frac{a}{1-4ta}w^2}$.
Plugging $t=\frac{1}{4}(\tau'-\tau)$, we obtain 
$$
I_{\tau}^{\tau'}(ce^{aw^2})=
\frac{c}{\sqrt{1{-}(\tau'{-}\tau)a}}
e^{\frac{a}{1{-}(\tau'{-}\tau)a}w^2}.
$$ 
To reveal its double-valued nature, we rewrite the above equality as follows:
\begin{equation}\label{2-to-2}
I_{\tau}^{\tau'}
(\frac{c}{\sqrt{1{-}\tau t}}e^{\frac{t}{1-\tau t}w^2})
=\frac{c}{\sqrt{1{-}\tau' t}}e^{\frac{t}{1-\tau' t}w^2}.
\end{equation} 
Since the branching singular point of 
the double-valued parallel section of 
the source space moves by the intertwiners, $I_{\tau}^{\tau'}$ 
must be viewed as a $2$-to-$2$ mapping.

\medskip
To describe \eqref{2-to-2} more clearlythe, we take two sheets with slit from
$\tau^{-1}$ to $\infty$, and denote points by $(t;+)_{\tau}$ or $(t;-)_{\tau}$.
$I_{\tau}^{\tau'}$ has the property that 
$I_{\tau}^{\tau'}((t;\pm)_{\tau})=(t;\pm)_{\tau'}$ as a set-to-set
mapping, and one may define this locally a 1-to-1 mapping. 
Note that 
$$
I_{\tau''}^{\tau}I_{\tau'}^{\tau''}I_{\tau}^{\tau'}((t,\pm)_{\tau})=(t,\pm)_{\tau},
$$ 
but this is neither the identity nor $-1$. This depends on $t$ discontinuously.

\bigskip
On the other hand, we want to retain the feature of complex one parameter group. 
For that purpose, we have to set ${:}e_*^{0w_*^2}{:}_{\tau}=1$ as the multiplicative unit 
for every expression. The problem is caused by another sheet, for we have to 
distinguish $1$ and $-1$.

It is important to recognize that 
there is no effective theory to understand such a vague 
system. This is something like an {\it air pocket} of the 
theory of point set topology.  
As it will be seen in the next section,  
this system forms an object which may be viewed 
as a {\it double covering group} of ${\mathbb C}$. 
This is absurd since ${\mathbb C}$ is simply 
connected !

\section{Extended notions for group-like objects}

Recall ${:}e_*^{tw_*^2}{:}_{\tau}$ does not form a group.  
However, using various expression parameters 
$\tau\not=0$, $e_*^{tw_*^2}$ behaves like a group.  
To handle the group-like nature of the 1-parameter family of 
$*$-exponential function $e_*^{tw_*^2}$, we introduce a notion of a
{\bf blurred covering group} of a topological group by 
using the notion of local groups.  
Consequently, $\{ e_*^{tw_*^2}; t\in{\mathbb C}\}$ is 
viewed as a blurred covering group of the abelian group 
$\{e^{tw^2}; t\in{\mathbb C}\}$. We need such a strange 
notion to understand the strange behaviour of the 
$*$-exponential functions for quadratic forms of several 
variables. 

In spite of Lie's third theorem which asserts that every finite
dimensional Lie algebra is the Lie algebra of a Lie group, 
we see in this section that the notion of local Lie groups 
is much wider than that of Lie groups, since it  
has to treat singular points.  

\noindent
{\bf A topological local group with unit}.\,\,
Recall that ${:}e_*^{tw_*^2}{:}_{\tau}$ is defined for 
$t\in {\mathbb C}{\setminus}\{\frac{1}{\tau}\}$. 
Abstracting the property of an open connected 
neighborhood $D$ of the identity $e$ of a topological group, we define 

\begin{definition}
A topological space $D$ is called a 
topological local group with the identity $e$, if the following 
conditions are satisfied:

\noindent
(a) For every $g\in D$, there is a 
neighborhoods $U$ of $g$ and $V$ of $e$ such that both $gh$ and $hg$ are defined continuously 
for every $g{\in}U$, $h{\in}V$.

\noindent
(b) $g^{-1}$ is defined on an open dense subset of $D$ and it is 
continuous.   

\noindent
(c) The associativity holds whenever they are defined.
\end{definition}

\subsection{A blurred covering group of a topological group}

Let $G$ be a locally simply arcwise connected topological 
group and let 
$\{{\mathcal O}_{\alpha}; \alpha\in I\}$ be 
an open covering of $G$.
It may be helpful to mind the correspondence as follows: 
$$ 
G\leftrightarrow {\mathbb C},\quad
\alpha\leftrightarrow \frac{1}{\tau},\quad
{\mathcal O}_{\alpha}\leftrightarrow
{\mathbb C}{\setminus}\{\frac{1}{\tau}\},\quad 
\varGamma\leftrightarrow{\mathbb Z}_2,\quad 
{\widetilde G}\leftrightarrow\{e_*^{tw_*^2}\},
$$ 
to understand the following abstract conditions: 

\medskip
\noindent
(a) For every $\alpha\in I$, ${\mathcal O}_{\alpha}$ contains 
the identity $e$. ${\mathcal O}_{\alpha}$ is called an 
{\bf abstract expression 

space}, and $\alpha$ is called an expression parameter.

\noindent
(b) For every $\alpha\in I$, ${\mathcal O}_{\alpha}$ is open, 
dense and connected, but it may not be simply connected.

\noindent
(c) For every $\alpha, \beta\in I$, there is a homeomorphism 
$\phi_{\alpha}^{\beta}: {\mathcal O}_{\alpha}\to {\mathcal O}_{\beta}$.

\noindent
(d) For every $g, h\in G$, there is $\alpha\in I$ and 
continuous path  $g(t), h(t)\in G$, $t\in [0,1]$,
such that 

$g(0)=h(0)=e$, $g(1)=g, h(1)=h$ and 
$g(t), h(t), g(t)h(t)$ are in ${\mathcal O}_{\alpha}$ 
for every $t\in [0,1]$.

\medskip
The open covering $\{{\mathcal O}_{\alpha}; \alpha\in I\}$ 
is called {\bf natural covering} of $G$ if it satisfies 
$(a){\sim}(d)$. The condition $(c)$ shows that there is an 
abstract topological space $X$ homeomorphic to every 
${\mathcal O}_{\alpha}$. 
We consider a connected covering space $\pi:\tilde X\to X$.    
This is same to say we consider a connected covering 
$\pi_{\alpha}: \widetilde{\mathcal O}_{\alpha}\to 
{\mathcal  O}_{\alpha}$ 
for each $\alpha$.
It is  easy see that $\pi_{\alpha}^{-1}(e)$ is a 
group given as a quotient group of the fundamental group 
of ${\mathcal O}_{\alpha}$. As $G$ is locally simply connected,
$\pi_{\alpha}^{-1}(e)$ forms a discrete group, and 
 $\phi_{\alpha}^{\beta}$ lifts to an isomorphism  
$\tilde\phi_{\alpha}^{\beta}: \pi^{-1}_{\alpha}(e)\to\pi^{-1}_{\beta}(e)$. 
We denote  $\pi^{-1}_{\alpha}(e)=\varGamma_{\alpha}$, and 
the isomorphism class is denoted by ${\varGamma}$.  

Choose $\tilde{e}_{\alpha}\in\pi_{\alpha}^{-1}(e)$ and 
call $\tilde{e}_{\alpha}$ a tentative identity. For any continuous
path $g(t)$ in ${\mathcal O}_{\alpha}$ such that $g(0)=g(1)=e$, 
the continuous chasing among the set $\pi^{-1}(g(t))$ starting at 
$\tilde{e}_{\alpha}$ gives a group element 
$\gamma \in \varGamma_{\alpha}$.
  
By a standard argument, it is easy to make 
$\widetilde{\mathcal O}_{\alpha}$ a local group 
such that $\pi_{\alpha}$ is a homomorphism: 
We define first that 
$\tilde e_{\alpha}\tilde e_{\alpha}=\tilde e_{\alpha}$.
For paths
$g(t), h(t), g(t)h(t)$ such that they are in ${\mathcal O}_{\alpha}$ for every 
$t\in [0,1]$ and $g(0)=h(0)=e$, 
we define the product 
by a continuous chasing among the set-to-set mapping 
$$
\pi^{-1}_{\alpha}(g(t))\pi^{-1}_{\alpha}(h(t))=\pi^{-1}_{\alpha}(g(t)h(t)).
$$
We set 
${\mathcal O}_{\alpha\beta}=
{\mathcal O}_{\alpha}\cap {\mathcal O}_{\beta},\quad 
{\mathcal O}_{\alpha\beta\gamma}=
{\mathcal O}_{\alpha}\cap {\mathcal O}_{\beta}\cap 
{\mathcal O}_{\gamma}$ for simplicity.

As $G$ is locally simply connected, the full inverse 
$\pi_{\alpha}^{-1}V$ of a simply connected neighborhood 
$V\subset{\mathcal O}_{\alpha}$ of the identity 
$e\in G$ is the disjoint union 
$\coprod_{\lambda}{\tilde V}_{\lambda}$, each member 
${\tilde X}_{\lambda}$ of which is homeomorphic to $V$.     

Moreover $\pi_{\alpha}^{-1}{\mathcal O}_{\alpha\beta}$ is 
also a local group for every $\beta$.

\subsubsection{Isomorphisms modulo $\varGamma$} 

For every $\alpha,\beta$, we define the notion of 
``isomorphism'' $I_{\alpha}^{\beta}$ 
of local groups, which corresponds to the notion of 
intertwiners in the previous section:  
$$
\begin{matrix}
\widetilde{\mathcal O}_{\alpha}&\supset&
\pi_{\alpha}^{-1}{\mathcal O}_{\alpha\beta}&
\overset{I_{\alpha}^{\beta}}\longrightarrow&
\pi_{\beta}^{-1}{\mathcal O}_{\beta\alpha}&\subset&
\widetilde{\mathcal O}_{\alpha}\\
\downarrow\pi_{\alpha}&{ }&{ }&{ }&{ }&{ }&
\downarrow\pi_{\beta}\\
{\mathcal O}_{\alpha}&\supset&
{\mathcal O}_{\alpha\beta}&
=\!=\!=&
{\mathcal O}_{\beta\alpha}&\subset&{\mathcal O}_{\beta}
\end{matrix}
$$
such that $I_{\beta}^{\alpha}=(I_{\alpha}^{\beta})^{-1}$, but 
the cocycle condition 
$I_{\alpha}^{\beta}I_{\beta}^{\gamma}I_{\gamma}^{\alpha}=1$
is not required for ${\mathcal O}_{\alpha\beta\gamma}$. 

Since the correspondence $I_{\alpha}^{\beta}$ 
does not make sense as a point set mapping, we should 
be careful for the definition.

Note that $I_{\alpha}^{\beta}$ is a collection of $1$-to-$1$ mapping 
$I_{\alpha}^{\beta}(g): \pi_{\alpha}^{-1}(g)\to
\pi_{\beta}^{-1}(g)$ for every 
$g\in{\mathcal O}_{\alpha\beta}=
{\mathcal O}_{\beta\alpha}$, which may not be continuous in 
$g$.   
  
For each $g$ there is a neighborhood 
$V_g$ of the identity $e$ such that 
$V_gg\subset {\mathcal O}_{\alpha\beta}$ and  the local trivialization  
$\pi_{\alpha}^{-1}(V_gg)=V_gg{\times}\pi_{\alpha}^{-1}(g)$.  
Thus $I_{\alpha}^{\beta}(g)$ extends to 
the correspondence  
$$
{\tilde I}_{\alpha}^{\beta}(h,g): 
\pi_{\alpha}^{-1}(hg)\to
\pi_{\beta}^{-1}(hg),\quad h\in V_g 
$$
which commutes with the local deck transformations. 

\begin{definition}
The collection $I_{\alpha}^{\beta}{=}\{I_{\alpha}^{\beta}(g); g\in{\mathcal O}_{\alpha\beta}\}$ 
is called an isomorphism 
modulo $\varGamma$, if 
$I_{\beta}^{\alpha}(hg){\tilde I}_{\alpha}^{\beta}(h,g)$ 
is in the group $\varGamma$ for every 
$g\in {\mathcal O}_{\alpha\beta}$ and $h\in V_g$. $($It follows the
continuity of $I_{\alpha}^{\beta}(hg)$ w.r.t. $h$.$)$
\end{definition}
The condition given by this definition means roughly 
that $I_{\alpha}^{\beta}(g)$ has discontinuity in $g$ only 
in the group $\varGamma$.

\medskip
${\widetilde G}=\{\widetilde{\mathcal O}_{\alpha}, 
\pi_{\alpha}, I_{\alpha}^{\beta}; \alpha,\beta\in I\}$ 
is called a {\bf blurred covering group} of $G$ 
if each $\widetilde{\mathcal O}_{\alpha}$ is a covering local 
group of ${\mathcal O}_{\alpha}$, where
$\{{\mathcal O}_{\alpha};\alpha\in I\}$
is a natural open covering of a locally simply arcwise 
connected topological group $G$ and   
$I_{\alpha}^{\beta}$ are isomorphisms 
modulo $\varGamma$. 

\medskip
Because of the failure of the cocycle condition, this 
object does neither form a covering group, nor a topological 
point set. However, this object looks like a covering group. 

\medskip 
For $g$, let $I_g$ be the set of expression parameters 
involving $g$; 
$I_g=\{\alpha\in I; {\mathcal O}_{\alpha}\ni g\}$. 
For every  $\alpha\in I(g,h,gh)=I_g\cap I_h\cap I_{gh}$,  
we easily see that 
$\pi_{\alpha}^{-1}(g)\pi_{\alpha}^{-1}(h)=
\pi_{\alpha}^{-1}(gh)$. In general, this is viewed as 
set-to-set correspondence, but if $g$ or $h$ is in a 
small neighborhood of the identity, we can make 
these correspondence a genuine point set mapping. 
Hence, we have the notion of indefinite small action 
or ``infinitesimal left/right action''  
of small elements to the object. This corresponds to  
the infinitesimal action 
$w_*^2{*}$ or ${*}w_*^2$ in the previous section.

\medskip
Next, we choose an element 
$\tilde{e}_{\alpha}\in\pi_{\alpha}^{-1}(e)$, and call it  
a local identity. On the other hand,  
$\pi_{\alpha}^{-1}(e)$ is called {\it the set of local identities} 
of ${\widetilde{G}}$. 
The failure of the cocycle condition gives that 
${\mathfrak M}_{\alpha}\tilde{e}_{\alpha}$ may not be a 
single point set, but forms a discrete abelian group. 
Hence an identity of our object is always a 
{\it local identity}.

Since $G$ is a locally simply connected, there is an open 
simply connected neighborhood $V_{\beta}$ of $e$ contained 
in ${\mathcal O}_{\beta}$. Hence, there is the unique lift 
$\tilde{V}_{\beta}$ through $\tilde{e}_{\beta}$. 
Setting 
$\tilde{V}_{\beta\gamma}=
\tilde{V}_{\beta}\cap\tilde{V}_{\gamma}$ e.t.c.,
 we see easily 
$I_{\beta}^{\gamma}(\tilde{V}_{\beta\gamma})
=\tilde{V}_{\gamma\beta}.$

\medskip
The $\{{\tilde g}_{\alpha}\in \tilde{\mathcal O}_{\alpha};\alpha\in I\}$
may be viewed as an element of ${\widetilde{G}}$ if 
$I_{\alpha}^{\beta}{\tilde g}_{\alpha}={\tilde g}_{\beta}$, 
but this is not a single point set by the same reason. 
In spite of this, one can distinguish individual points within 
a small local area.

\medskip
The ${*}$-exponential function $e_*^{zw_*^2}$ may be viewed as a 
blurred covering group of ${\mathbb C}$ by treating this as a family 
$\{{:}e_*^{zw_*^2}{:}_{\tau}; \tau \}$, where the feature of complex 
one parameter group is retained. 

\subsection{Several remarks for the equation $(w_*^2{-}a^2){*}f=0$.}\label{S140}
If $f$ satisfies $w_*^2{*}f=a^2f$, then $e^{ta^2}f$ is the real analytic
solution of the evolution equation
$\frac{d}{dt}f_t(w)=w_*^2{*}f_t(w)$ with the initial value $f$.  
Hence, one may write $e_*^{tw_*^2}{*}f=e^{ta^2}f$ by defining the
$*$-product by this way. Next one gives a justification: 
\begin{prop}\label{dameresidue}
If ${\rm{Re}}\,\tau>0$, then 
${:}e_*^{tw_*^2}{*}\delta_*(w+\alpha){:}_\tau$  
is holomorphic in $t\in{\mathbb C}$. That is,
$\{{:}e_*^{tw_*^2}{:}_{\tau}; t\in{\mathbb C}\}$  
acts on $\delta_*(w+\alpha){:}_\tau$ as a genuine 
one parameter group. That is, 
${:}e_*^{tw_*^2}{:}_{\tau}{*_\tau}
{:}\delta_*(w{+}\alpha){:}_{\tau}=
e^{t\alpha^2}{:}\delta_*(w{+}\alpha){:}_{\tau}$.
$($Cf.\eqref{nicerep}.$)$
\end{prop}

\noindent
{\bf Proof}\,\,\,Since $f(w){*_\tau}e^{aw}=
f(w{+}\frac{a\tau}{2})e^{aw}$, we see 
$$
{:}e_*^{tw_*^2}{:}_{\tau}{*_\tau}
{:}e_*^{i\sigma(w{+}\alpha)}{:}_\tau
=\frac{1}{\sqrt{1{-}t\tau}}
e^{\frac{t}{1{-}t\tau}w^2
{+}\frac{i\sigma}{1-t\tau}w
{+}i\sigma\alpha
{-}\frac{\tau}{4(1{-}t\tau)}\sigma^2}.
$$
If ${\rm{Re}}\,\tau>0$ and $t\not=\tau^{-1}$, the integral
$$
\int_{\mathbb R}
{:}e_*^{tw_*^2}{:}_{\tau}{*_\tau}
{:}e_*^{i\sigma(w{+}\alpha)}{:}_\tau d\sigma 
=
\frac{1}{\sqrt{1{-}t\tau}}
e^{\frac{t}{1{-}t\tau}w^2{-}
\frac{1}{\tau(1{-}t\tau)}(w{+}\alpha(1{-}t\tau))^2}
\!\int_{\mathbb R}\!e^{-\frac{\tau}{4(1{-}t\tau)}
(\sigma{-}\frac{2i}{\tau}(w{+}\alpha(1{-}t\tau))^2}d\sigma  
$$
converges. By the similar calculation as in \eqref{eq:Fouriertrsf} 
giveds 
$$
\!\int_{\mathbb R}\!e^{-\frac{\tau}{4(1{-}t\tau)}
(\sigma{-}\frac{2i}{\tau}(w{+}\alpha(1{-}t\tau))^2}d\sigma
=
\frac{2\sqrt{\pi(1{-}t\tau)}}{\sqrt{\tau}}
$$ 
Note that $t=\tau^{-1}$ is a removable singularity in this integral.
Hence,  
$$
{:}e_*^{tw_*^2}{:}_{\tau}{*_\tau}
{:}\delta_*(w{+}\alpha){:}_{\tau}=
\frac{1}{\sqrt{\pi\tau}}
e^{\alpha^2 t}e^{-\frac{1}{\tau}(w{+}\alpha)^2}=
e^{t\alpha^2}{:}\delta_*(w{+}\alpha){:}_{\tau}=
e^{t(-\alpha)^2}{:}\delta_*({-}\alpha{-}w){:}_{\tau}
  \qquad \qquad\qquad \qquad\qquad \Box
$$ 

\medskip
\noindent
{\bf Note}\,\, This gives an example that 
even though the family $\{e_*^{tw_*^2}, t{\in}\mathbb C\}$ 
does not form a genuine group, this can act as a genuine 
one parameter group on some restricted family. This gives also 
an example that the formula 
$e_*^{tw_*^2}=\int e^{tx^2}\delta_*(x{-}w)dx$ does not extend for 
$t\in {\mathbb C}$. 

\medskip
Note the equation $({\alpha}^2{-}w_*^2){*}f=0$ can be solved by the
Fourier transform. Namely, by setting 
$f={f}_{{\alpha}}(w)=\int{\hat f}_{\alpha}(t)e_*^{itw}dt$, the equation is changed into 
$$
\int{\hat f}_{{\alpha}}(t)({\alpha}^2{-}w_*^2){*}e_*^{itw}dt=
\int{\hat f}_{{\alpha}}(t)({\alpha}^2{+}\frac{d^2}{dt^2}e_*^{itw})dt=0.
$$
Integration by parts gives that  
${\hat f}_{{\alpha}}(t)
=ae^{i\alpha\,t}+be^{-i\alpha\,t},\quad 
a,\,b\in{\mathbb C}.$ 
Hence we have 
$$
{f}_{{\alpha}}(w)=
\int(ae^{i\alpha t}+be^{-i\alpha t})e_*^{itw}dt 
\quad a,\,b\in{\mathbb C}.
$$
If ${\rm{Re}}\,\tau {>}0$, then the r.h.s makes sense for any 
${\alpha}\in \mathbb C$ to give the solution. This is equivalent to give 
the solution as 
$$
{f}_{{\alpha}}(w)=a\delta_*(w{+}\alpha){+}b\delta_*(w{-}\alpha).
$$
By \eqref{eq:Fouriertrsf}, the $\tau$-expression of 
${f}_{{\alpha}}(w)$ is given by 
$$
{:}{f}_{{\alpha}}(w){:}_{\tau}=
\frac{1}{\sqrt{\pi\tau}}
\big(ae^{-\frac{1}{\tau}(w{+}{\alpha})^2}{+}
   be^{-\frac{1}{\tau}(w{-}{\alpha})^2}\big).
$$ 
Thus, the equation $({\alpha}^2{-}w_*^2){*}f=0$ is solved uniquely 
by the boundary data $f_{{\alpha}}(0)$ and $f'_{{\alpha}}(0)$.
Let $\Phi_{{\alpha}}(w,\tau)$, $\Psi_{{\alpha}}(w,\tau)$ be the 
solutions of $({\alpha}^2{-}w_*^2){*}f=0$ such that 
$$
\Phi_{{\alpha}}(0,\tau)=1,\,\,\Phi'_{{\alpha}}(0,\tau)=0,\quad 
\Psi_{{\alpha}}(0,\tau)=0,\,\,\Psi'_{{\alpha}}(0,\tau)=1.
$$
As these are linear combinations of $*$-delta functions, 
Proposition\,\ref{dameresidue} shows that 
$e_*^{zw_*^2}{*}\Phi_{{\alpha}}(w,*)$, $e_*^{zw_*^2}{*}\Psi_{{\alpha}}(w,*)$
are defined without singularity.  
This is a phenomenon that the singular point of the 
differential equation 
$\frac{d}{dt}f_t=(w^2+\frac{\tau}{2}){*_\tau}f_t$
depends on initial functions. If $f_0=1$ then the 
solution ${:}e_*^{tw_*^2}{:}_{\tau}$ has a singular point 
at $t=\tau^{-1}$, but if $f_0=\Phi_{\nu}(w,\tau)$ or 
$\Psi_{\nu}(w,\tau)$, then there is no singular point.

\medskip
On the other hand, the integral along a closed path   
$\int_{C^2}e_*^{z(\nu{+}w_*^2)}dz$ satisfies 
$(\nu{+}w_*^2){*}\int_{C^2}e_*^{z(\nu{+}w_*^2)}dz=0$ 
where $C^2$ is the path turning around the same circle 
$C$ twice avoiding singular point so that integrand 
is closed on that path. 
As $\int_{C^2}e_*^{z(\nu{+}w_*^2)}dz$ is a function of $w^2$,
we see 
$\int_{C^2}{:}e_*^{z(\nu{+}w_*^2)}{:}_{\tau}dz
{=}\alpha\Phi_{\nu}(w,\tau)$ 
and the constant $\alpha$ is given by the value at $w^2=0$. 
Hence, we have 
\begin{equation}\label{sqrtform}
\int_{C^2}{:}e_*^{z(\nu{+}w_*^2)}{:}_{\tau}dz=
\int_{C^2}\frac{e^{z\nu}}{\sqrt{1{-}z\tau}}dz\,
\Phi_{\nu}(w,\tau).
\end{equation} 
Computing the Laurent expansion of
$\frac{e^{(\tau^{-1}+s^2)\nu}}{s\sqrt{-\tau}}$ 
at $s{=}0$ and setting $z{=}s^2$ we see 
$\int_{C^2}\frac{e^{z\nu}}{\sqrt{1{-}z\tau}}dz=0$ 
by the fact that the secondary residue $a_{-2}$ does not 
appear in the Laurent series.
Hence, we have the following extraordinary property:
\begin{prop}\label{ResRes00}
$\int_{C^2}e_*^{z(\nu{+}w_*^2)}dz=0$ for any closed path $C^2$. 
\end{prop}

Besides integrals along closed path $C$, the integral along a
non-compact path $\Gamma$: 
$$
\int_{\Gamma}{:}e_*^{z(\nu{+}w_*^2)}{:}_{\tau}dz=
\int_{\Gamma}\frac{e^{z\nu}}
{\sqrt{1{-}z\tau}}
e^{\frac{z}{1{-}z\tau}w^2}dz 
$$
converges if $\Gamma$ is suitably choosed  
under  ${\rm{Re}}\,\nu>0$. By the continuity of 
$(\nu{+}w_*^2)*$, the integral must satisfy  
$$
(\nu{+}w_*^2)*\int_{\Gamma}e_*^{z(\nu{+}w_*^2)}dz
=\int_{\Gamma}\frac{d}{dz}e_*^{z(\nu{+}w_*^2)}dz=0.
$$
This integral has a remarkable feature that this is given as the 
difference of two inverses of $\nu{+}w_*^2$: Let $\Gamma_{\pm}$ 
be two different paths from ${-}\infty$ to $0$ such that 
$\Gamma{=}\Gamma_+\setminus \Gamma_-$.  
Then,  
$\int^0_{\Gamma_+}e_*^{z(\nu{+}w_*^2)}dz{-}\int^0_{\Gamma_-}e_*^{z(\nu{+}w_*^2)}dz$
is nontrivial and satisfies the equation 
$(\nu{+}w_*^2){*}f=0$.

\subsection{Residues and Laurent series} 

Note that ${:}e_*^{zw_*^2}{:}_{\tau}$ has a branching singular point at 
$z=\tau^{-1}$. Let $D$ be a small disk with the center at $\tau^{-1}$. 
Let $s$ be the complex coordinate of the double covering space $\tilde{D}_*$ 
of $D{\setminus}\{\frac{1}{\tau}\}$ such that 
$z=s^2{+}\tau^{-1}$. ${:}e_*^{zw_*^2}{:}_{\tau}$ is viewed 
as a single valued holomorphic function of $s$ on the double 
covering space $\tilde{D}_*$. 
The residue at $s=0$ is defined as the coefficient 
$a_{-1}$ of $1/s$ of the Laurent-series expansion 
at the isolated singular point $s=0$. We extend the term {\it residue} 
to be $0$ at a regular point.

\medskip
Using \eqref{eq:expquad}, we see that the 1-form  
\begin{equation}\label{negativeodd}
{:}e_*^{(\tau^{-1}{+}s^2)w_*^2}{:}_{\tau}ds=
\frac{ds}{s}e^{-\frac{w^2}{\tau^2s^2}}\frac{1}{\sqrt{-\tau}}e^{-\frac{1}{\tau}w^2}
=\frac{1}{\sqrt{-\tau}}e^{-\frac{1}{\tau}w^2}
\big(\frac{1}{s}-\frac{w^2}{\tau^2s^3}{+}\frac{w^4}{2!\tau^4s^5}{-}\cdots\big)ds
\end{equation}
has terms only of negative odd degrees w.r.t. $s$.  
The 2-form ${:}e_*^{(\tau^{-1}{+}s^2)w_*^2}{:}_{\tau}ds$ may be written as 
${:}e_*^{zw_*^2}{:}_{\tau}\frac{dz}{2\sqrt{z-\tau^{-1}}}$ by setting a suitable slit.
The Cauchy's integral theorem gives that the residue is given by    
given by 
\begin{equation}\label{Reswsquar}
{\rm{Res}}_{z=\tau^{-1}}({:}e_*^{zw_*^2}{:}_{\tau})
=\frac{1}{2\pi i}\int_{\tilde{C}}
{:}e_*^{(\tau^{-1}{+}s^2)w_*^2}{:}_{\tau}ds
=\frac{1}{\sqrt{-\tau}}e^{-\frac{1}{\tau}w^2}
\frac{1}{2\pi i}
\int_{\tilde C}\frac{1}{s}e^{-\frac{1}{s^2\tau^2}w^2}ds
=\frac{1}{\sqrt{-\tau}}e^{-\frac{1}{\tau}w^2}
\end{equation}
where $\tilde{C}$ corresponds $C^2$ the path turning around  
the same circle $C=\partial D$ twice so that the path 
is closed. As there are only two singular points $s=0$ and 
$s=\infty$, one needs not to take the radius of $C$ small, 
but one may set $|s|=1$.  
It is very suggestive to compare the residue formula with the 
 \eqref{eq:Fouriertrsf}. If ${\rm{Re}}\,\tau>0$, 
then $\frac{1}{\sqrt{-\tau}}e^{-\frac{1}{\tau}w^2}
{=}\sqrt{-\pi}{:}\delta_*(w){:}_{\tau}$. 
Note also that the integral obtaining the residue  may be 
replaced as follows by taking the $\pm$ sheet and the slit in 
mind:  
\begin{equation}\label{Reswsquar22}
{\rm{Res}}_{z=\tau^{-1}}({:}e_*^{z(\nu{+}w_*^2)}{:}_{\tau})=
\frac{1}{2\pi i}\int_{C^2}
{:}e_*^{z(\nu{+}w_*^2)}{:}_{\tau}\frac{dz}{2\sqrt{z{-}\tau^{-1}}}
=
\frac{1}{2\pi i}\int_{C}
{:}e_*^{zw_*^2}{:}_{\tau}\frac{e^{z\nu}dz}{\sqrt{z{-}\tau^{-1}}}
\end{equation}
where $C^2$ means the union $C_+$ and $C_-$ of $C$ viewed as 
a curve in $\pm$-sheets. Note that the $\pm$-sign changes 
on $\pm$ sheets. The existence of the slit keeps the 
integrand single value, and $dz$ is treated $-dz$ 
in the negative sheet. 
Hence $\frac{dz}{\sqrt{z{-}\tau^{-1}}}$ does not change sign 
on the opposite sheet.
%
%

\subsubsection{Discontinuity of Laurent coefficients} 

Recall that 
\begin{equation}\label{L-series}
{:}e_*^{(\tau^{-1}{+}s^2)(\nu{+}w_*^2)}{:}_{\tau}=
e^{\tau^{-1}\nu}
\frac{1}{\sqrt{-\tau}}e^{{-}\frac{1}{\tau}w^2}
\,\,\frac{1}{s}e^{\nu s^2{-}\frac{1}{s^2}\frac{w^2}{\tau^2}}.
\end{equation}
We have the Laurent series for 
$\frac{1}{s}e^{\nu s^2{-}\frac{1}{\tau^2s^2}w^2}$ as  
$$
\cdots{+}\frac{c_{-(2k{+}1)}(\nu,\tau,w)}{s^{2k{+}1}}
{+}\cdots{+}\frac{c_{-1}(\nu,\tau)}{s}{+}c_1(\nu,\tau,w)s{+}
c_{3}(\nu,\tau,w)s^3{+}\cdots
$$
without terms of even degree.  
We have $c_{2k+1}(\nu,\tau,w)=0$ at $\nu=0$ for $k\geq 0$ by \eqref{negativeodd}.
Hence the Laurent series of 
${:}e_*^{(\tau^{-1}{+}s^2)(\nu{+}w_*^2)}{:}_{\tau}$ is given by 
\begin{equation}\label{akak}
\begin{aligned}
\sum_{k\in\mathbb Z}&a_{2k{-}1}(\nu,\tau,w)s^{2k{-}1}=
e^{\tau^{-1}\nu}
\frac{1}{\sqrt{-\tau}}
e^{{-}\frac{1}{\tau}w^2}\sum_{k}c_{2k-1}(\nu,\tau,w)s^{2k{-}1}, \\
a_{2k{-}1}(\nu,\tau,w)&= 
{\rm{Res}}_{s=0}({:}s^{-2k}e_*^{(\tau^{-1}{+}s^2)(\nu{+}w_*^2)}{:}_{\tau}),\quad
a_{{-}1}(\nu,\tau,w)= 
\frac{e^{\frac{\nu}{\tau}}}{\sqrt{-\tau}}
e^{-\frac{1}{\tau}w^2}
\sum_k\frac{(-\nu)^k}{k!k!}(\frac{w}{\tau})^{2k}.
\end{aligned}
\end{equation}
Note that every  $a_{2k-1}(\nu,\tau,w)$ is written in the form 
$$
a_{2k-1}(\nu,\tau, w)=e^{\tau^{-1}\nu} 
\frac{1}{\sqrt{-\tau}}e^{{-}\frac{1}{\tau}w^2}
p_{2k{-}1}(\tau^{-1},w^2)
$$
by using a certain polynomial $p_{2k{-}1}(\tau^{-1},w^2)$. The following is easy to see: 
\begin{prop}\label{Laurentcoeff}
$a_{2k-1}(0,\tau, w)=0$ for $2k-1\geq 0$, and 
$a_{2k-1}(\nu,\tau,0)=0$ for $2k{-}1\leq -2$. Hence 
$a_{2k-1}(0,\tau,0)=0$ except for $k=0$: $a_{-1}(0,\tau,0)=\frac{1}{\sqrt{-\tau}}$. 
\end{prop}

A strange fact arises by writing these as integrals:  
$$
a_{2k{-}1}=\frac{1}{2\pi i}\int_{\tilde C}
{:}s^{-2k}e_*^{(\tau^{-1}{+}s^2)(\nu{+}w_*^2)}{:}_{\tau}ds
=\frac{1}{\sqrt{-\tau}}e^{{-}\frac{1}{\tau}w^2}
e^{\frac{\nu}{\tau}}\frac{1}{2\pi i}\int_{\tilde C} 
\frac{1}{s^{2k{+}1}}e^{\nu s^2{-}\frac{1}{\tau^2s^2}w^2}ds
$$  
where $\tilde C$ is any simple closed curve in the covering space 
${\mathbb C}{\setminus}\{\tau^{-1}\}$ 
turning positively around $\tau^{-1}$. 
By Cauchy's theorem, it does not depend on $\tilde C$, hence it may 
be infinitesimally small.
Integration by parts gives  
\begin{equation}\label{nottrick}
\begin{aligned}
{:}(\nu{+}w_*^2){:}_{\tau}{*_{\tau}}a_{2k{-}1}
=&{:}\frac{1}{2\pi i}\int_{\tilde{C}}
\frac{1}{2}s^{{-}2k{-}1}\frac{d}{ds}
e_*^{(\tau^{-1}{+}s^2)(\nu{+}w_*^2)}ds{:}_{\tau}\\
=&(k{+}1/2)\frac{1}{2\pi i}\int_{\tilde{C}}
s^{-2k{-}2}
{:}e_*^{(\tau^{-1}{+}s^2)(\nu{+}w_*^2)}{:}_{\tau}ds
=(k{+}1/2)a_{2k{+}1}.
\end{aligned}
\end{equation}
( If $\nu=0$,\, \eqref{negativeodd} shows that $a_{2k{+}1}=0$ for
$k\geq 0$.)

\medskip 
There is a strange phenomenon as follows:
\begin{prop}\label{FundRes00} In spite that \eqref{nottrick} implies 
${:}(\nu{+}w_*^2){:}_{\tau}{*_{\tau}}a_{2k{-}1}(\nu,\tau)\not=0$, we have 
${:}e_*^{t(\nu{+}w_*^2)}{:}_{\tau}{*_\tau}a_{2k{-}1}(\nu,\tau)=0$  
for any $t\not=0$, and this is not continuous at $t=0$. 
Hence, differentiating by $t$ at $t=0$ is prohibited. 
\end{prop} 

\noindent
{\bf Proof}\,\,Using the formula \eqref{Reswsquar} and 
the exponential law, we have 
$$
{:}e_*^{t(\nu{+}w_*^2)}{:}_{\tau}{*_{\tau}}
\frac{1}{2\pi i}\int_{\tilde{C}}s^{-2k}
{:}e_*^{(\tau^{-1}{+}s^2)(\nu{+}w_*^2)}ds
=
\frac{1}{2\pi i}\int_{\tilde{C}}s^{-2k}
{:}e_*^{(t{+}\tau^{-1}{+}s^2)(\nu{+}w_*^2)}{:}_{\tau}ds.
$$
This is ensured since both sides satisfies the same 
differential equation 
$$
\frac{d}{dt}f_t=(\nu{+}w_*^2){*}f_t,\quad 
f_0=\frac{1}{2\pi i}\int_{\tilde{C}}s^{-2k}
{:}e_*^{(\tau^{-1}{+}s^2)(\nu{+}w_*^2)}ds.
$$
Note the radius of ${\tilde C}$ can be infinitesimally 
small by virtue of Cauchy's integral theorem. 
Hence if $t\not=0$, then $t{+}\tau^{-1}$ is 
outside the path of integration. Thus it must vanish. 
${}$ \hfill $\Box$ 

\bigskip
Apparently, this is caused that $\tilde C$ is chosen infinitesimally
small. Therefore, if $\tilde C$ is big enough, then the integral 
$\frac{1}{2\pi i}\int_{\tilde{C}}s^{-2k}
{:}e_*^{(a{+}\tau^{-1}{+}s^2)(\nu{+}w_*^2)}{:}_{\tau}ds$
is defined to gives $a_{2k-1}$.  
Thus, to avoid possible confusion,  it is better to fix the definition of the residue by 
\begin{equation}\label{defresres}
{\rm{Res}}_{s=0}f(s)=\lim_{r\to 0}\int_{C(r)}f(s)ds
\end{equation}
where $C(r)$ is a circle of radius $r$ with the center at $s=0$.

\medskip 

Although Proposition\,\ref{FundRes00} shows 
${:}(\nu{+}w_*^2){:}_{\tau}{*_\tau} 
{\rm{Res}}_{z=\tau^{-1}}({:}e_*^{z(\nu{+}w_*^2)}{:}_{\tau})\not=0$ 
in general, the case $\nu=0$ is rather special. 
By \eqref{negativeodd}, we see that 
${:}w_*^2{:}_{\tau}{*_{\tau}}
{\rm{Res}}_{z=\tau^{-1}}({:}e_*^{zw_*^2}{:}_{\tau})=0$.
Hence, there must be a constant $\alpha$ such that 
$$
{\rm{Res}}_{z=\tau^{-1}}({:}e_*^{zw_*^2}{:}_{\tau})
{=}\alpha\Phi_{0}(w^2,\tau)
$$ 
where $\alpha$ is given by the value at $w^2=0$. 
Hence,  we have an equality 
\begin{equation}\label{residue}
{\rm{Res}}_{z=\tau^{-1}}({:}e_*^{zw_*^2}{:}_{\tau})
=
{\rm{Res}}_{z=\tau^{-1}}(\frac{1}{\sqrt{1{-}z\tau}})
\Phi_{0}(w^2,\tau)
=
\frac{1}{\sqrt{-\tau}}\Phi_{0}(w^2,\tau).
\end{equation}

This is strange, for the r.h.s of \eqref{residue} satisfies 
${:}e_*^{tw_*^2}{:}_{\tau}{*_{\tau}}
\frac{1}{\sqrt{-\tau}}\Phi_{0}(w^2,\tau)=
\frac{1}{\sqrt{-\tau}}\Phi_{0}(w^2,\tau),$
but Proposition\,\ref{FundRes00} shows 
${:}e_*^{tw_*^2}{:}_\tau {*_\tau}
{\rm{Res}}_{z=\tau^{-1}}({:}e_*^{zw_*^2}{:}_{\tau})=
{\rm{Res}}_{z=\tau^{-1}}({:}e_*^{(t{+}z)w_*^2}{:}_{\tau})=0$  
for $t\not=0$ by the computations as residues.    
Recall that $\Phi_{0}(w^2,\tau)$ is defined by 
the differential equation, while 
${\rm{Res}}_{z=\tau^{-1}}({:}e_*^{zw_*^2}{:}_{\tau})$ 
is defined by the integral on an infinitesimally small circuit. 
The equality 
${\rm{Res}}_{z=\tau^{-1}}({:}e_*^{zw_*^2}{:}_{\tau})
=
\frac{1}{\sqrt{-\tau}}\Phi_{0}(w^2,\tau)$ 
holds only on some restricted stage.

One of the way to avoid such a strange impression is to regard 
${\rm{Res}}_{z=\tau^{-1}}{:}e_*^{z(w_*^2{+}\nu)}{:}_{\tau}$ as a {\it formal distribution} 
supported only on the surface  $S_*$: $z=\tau^{-1}$. 

To treat ``functions'' such as residues,   
it is convenient to use the notion of formal distributions.  
This is the notion based on the calculations of 
residues by regarding Laurent polynomials as ``test functions''.
Formal distributions are used extensively in conformal field theory.

\subsubsection{Covariant differentials and ${*}$-product integrals} 

Note in general, the Laurent coefficient $a_{2k-1}$ of  
${:}e_*^{(z{+}s^2)(\nu{+}w_*^2)}{:}_{\tau}$ is obtained in the formula  
$$
{\rm{Res}}_{s=0}({:}s^{-2k}e_*^{(z{+}s^2)(\nu{+}w_*^2)}{:}_{\tau})=
\left\{
\begin{matrix}
\medskip
a_{2k-1} &z=\tau^{-1}\\
0  & z\not=\tau^{-1}
\end{matrix}
\right.
$$
This is a formal distribution of $(z,\tau){\in}{\mathbb C}_*^2$. We
denote this by $R_{2k{-}1}(z,\tau)$, i.e.   
$$
R_{2k{-}1}(z,\tau)=
{\rm{Res}}_{s=0}(s^{-2k}{:}e_*^{(\tau^{-1}{+}s^2)w_*^2}{:}_{\tau})\delta(z{-}\tau^{-1}).
$$

If we set 
$E(z,\tau)(s)={:}e_*^{(z{+}s^2)(\nu{+}w_*^2)}{:}_{\tau},\,\, s\not=0,$
and regard this a formal distribution supported on $z=\tau^{-1}$, 
then Laurent expansion theorem shows 
$$
E(z,\tau)(s)=\sum_{k\in{\mathbb Z}}R_{2k{-}1}(z,\tau)s^{2k{-}1},\quad 0<|s|<\infty.
$$

\medskip
Now we are interested only in the function 
${R}_{2k{-}1}(\tau^{-1},\tau,)$ restricted 
in the surface $S_*$. 
Note that the infinitesimal intertwiner is given by  
$\lim_{\delta\to 0}I_{z^{-1}}^{(z{+}\delta)^{-1}}
={-}\frac{1}{4z^2}\partial^2_{w}$
for every $H\!ol({\mathbb C})$-valued function$f(z,\tau, w)$. 
We now define  
\begin{equation}\label{covariant}
\nabla_{z}f(z,z^{-1},w)=\partial_{z}f(z,\tau,w)\big|_{\tau=z^{-1}}
=\partial_{z}(f(z,z^{-1},w))
{+}\frac{1}{4z^2}\partial_w^2f(z,z^{-1},w).
\end{equation}
This will be called {\it covariant} or {\it co-moving}
differentiation. In other words, we define 
\begin{equation}\label{covariant00}
\nabla_{\tau^{-1}}f(\tau^{-1},\tau,w)
=\lim_{\delta\to 0}\frac{1}{\delta}
\Big(I_{\tau}^{(\tau^{-1}{+}\delta)^{-1}}f(\tau^{-1}{+}\delta,\tau,w){-}f(\tau^{-1},\tau,w)\Big).
\end{equation} 
Noting that
$\partial_{\tau}f(z,\tau,w)={-}\frac{1}{4}\tau\partial^2_{w}$,  
we extend the notion of covariant derivative to functions
$f(z,\tau)$ without $w$ by 
$$
\nabla_{z}f(z,z^{-1})=\partial_{z}f(z,\tau)\big|_{\tau=z^{-1}}.
$$
We see easily for every pair of integers $(m,k)$
$\partial_{z}((m{+}k)z^m{-}m\tau^kz^{m+k})\big|_{\tau=z^{-1}}=0.$
Hence setting $f_{k,m}(z,\tau)=(m{+}k)z^m{-}m\tau^kz^{m+k}$, one
may treat this a {\it parallel polynomial of degree k} as 
$\nabla_{z}f_{k,m}(z,z^{-1})=0$.
However, we do not use $\partial_{z}(\log z{-}z\tau)\big|_{\tau=z^{-1}}=0$ for
$\log z$ is multi-valued. 
Such parallel polynomials forms a commutative algebra. We call these 
{\it parallel polynomials on} $z=\tau^{-1}$ and denote this by 
${\mathcal P}[S_*].$

\begin{prop}\label{diffeqevol}
Every Laurent coefficient $a_{2k-1}(\nu,w^2)(\tau)$ of   
${:}e_*^{(\tau^{-1}{+}s^2)(\nu{+}w_*^2)}{:}_{\tau}$ satisfies the 
differential equation 
\begin{equation}\label{nicerelation22}
\nabla_{\tau^{-1}}a_{2k-1}(\nu,w^2)(\tau)={:}(\nu{+}w_*^2){:}_{\tau}{*_{\tau}}a_{2k-1}(\nu,w^2)(\tau).
\end{equation}
\end{prop}

%
%
We insists that $\nabla_{\tau^{-1}}$ is the notion of co-moving
derivative. Ihe equality above may be written as   
\begin{equation}\label{musterequ}
\nabla_{\tau^{-1}}{:}e_*^{(\tau^{-1}{+}s^2)(\nu{+}w_*^2)}{:}_{\tau}=
{:}(\nu{+}w_*^2){:}_{\tau}{*_{\tau}}{:}e_*^{(\tau^{-1}{+}s^2)(\nu{+}w_*^2)}{:}_{\tau},\quad 
(s\not=0).
 \end{equation}

\subsubsection{Equation 
$\nabla_{\tau^{-1}}F(\tau^{-1},\tau){=}
{:}(\nu{+}w_*^2){:}_{\tau}{*_{\tau}}F(\tau^{-1},\tau)$} 

Note that for every parallel polynomial $c(z,\tau)$,  
$c(\tau^{-1},\tau)F(\tau^{-1},\tau)$ must satisfy the original
equation. 
Rewrite the equation $\nabla_{\tau^{-1}}F(\tau^{-1},\tau){=}
{:}(\nu{+}w_*^2){:}_{\tau}{*_{\tau}}F(\tau^{-1},\tau)$ by using 
 \eqref{covariant} in l.h.s and by using the product formula in
 r.h.s. Then, the highest parts are cancelled out and 
the equation becomes a differential equatoin of 1-st order: 
\begin{equation}\label{covariant11}
\partial_{\tau^{-1}}F(\tau^{-1},\tau)=
\tau w\partial_w F(\tau^{-1},\tau){+}(w^2{+}\nu{+}\frac{\tau}{2})F(\tau^{-1},\tau)
\end{equation}

Recalling that $F(\tau^{-1},\tau)$ involves the variable (generator) $w$,
we can solve \eqref{covariant11} by a standard manner. 
First set 
$F(\tau^{-1},\tau)=e^{-\tau^{-1}w^2}G(\tau^{-1},w)$. Then,
\eqref{covariant11} turns out 
$\partial_{\tau^{-1}}G=\tau w\partial_wG{+}(\nu{+}\frac{\tau}{2})G$.
 Thus, we have 
\begin{equation}\label{relativity}
F(\tau^{-1},\tau)=\sqrt{\tau^{-1}}e^{\tau^{-1}(\nu{-}w^2)}H(\tau^{-1}w).
\end{equation}
using  an arbitraly holomorphic function $H(z)$.
If the initial data is given at $\tau^{-1}=1$ and $F(1,1)=1$, then 
$$
F(\tau^{-1},\tau)=\sqrt{\tau^{-1}}e^{\tau^{-1}(\nu{-}w^2)}e^{-(\nu{-}\tau^{-2}w^2)}.
$$ 

\begin{prop}\label{nonsingular}
If the initial data is not singular, then there is no singular point 
on the solution of 
$$
\nabla_{\tau^{-1}}F(\tau^{-1},\tau){=}
{:}(\nu{+}w_*^2){:}_{\tau}{*_{\tau}}F(\tau^{-1},\tau).
$$
\end{prop}

\bigskip
On the other hand, there must be a holomorphic function $H(z,s)$ on
${\mathbb C}_*{\times}{\mathbb C}_*$ such that  
$$
{:}e_*^{(\tau^{-1}{+}s^2)(\nu{+}w_*^2)}{:}_{\tau}
=\sqrt{\tau^{-1}}e^{\tau^{-1}(\nu{-}w^2)}H(\tau^{-1}w, s).
$$ 
Putting $\tau^{-1}=-s^2$, we have 
$1={is}e^{-s^2(\nu{-}w^2)}H({-}s^{2}w,s)$ and then 
$H(-s^2w,s)=\frac{1}{is}e^{s^2(\nu{-}w^2)}$. Hence, 
$H(z,s)=\frac{1}{is}e^{\nu s^2{-}z^2s^{-2})}$, and 
$$
{:}e_*^{(\tau^{-1}{+}s^2)(\nu{+}w_*^2)}{:}_{\tau}=
\frac{1}{\sqrt{\tau}}e^{\frac{1}{\tau}(\nu{-}w^2)}
\frac{1}{is}e^{(\nu s^2{-}\frac{1}{\tau^2 s^2}w^2)}.
$$
This is nothing but the $\tau$-expression of 
$e_*^{(\tau^{-1}{+}s^2)(\nu{+}w_*^2)}$. 

\section{Isolated singular points and formal distributions}
In this section, we treat 
$E(z,\tau){=}{\rm{Res}}_{s=0}{:}e_*^{(z{+}s^2)(w_*^2{+}\nu)}{:}_{\tau}$ 
as a formal distribution. 
Recall that 
$$
{:}e_*^{(\tau^{-1}{+}s^2)(w_*^2{+}\nu)}{:}_{\tau}=
e^{\tau^{-1}\nu}
\frac{1}{\sqrt{-\tau}}e^{{-}\frac{1}{\tau}w^2}
\,\,\frac{1}{s}e^{\nu s^2{-}\frac{1}{s^2}\frac{w^2}{\tau^2}}.
$$

For every Laurent polynomial $f(s)\in {\mathbb C}[s,s^{-1}]$, we set 
$$
\{f(s)\}=f(s){:}e_*^{(\tau^{-1}{+}s^2)(w_*^2{+}\nu)}{:}_{\tau}
\in {\mathbb C}[s,s^{-1}]{:}e_*^{(\tau^{-1}{+}s^2)(w_*^2{+}\nu)}{:}_{\tau}.
$$
Note that $\{f(s){+}g(s)\}$ and $\{f(s)g(s)\}$ are defined as
usual. Moreover, we see by definition 
$$
f(s)\{g(s)\}=\{f(s)g(s)\}.
$$
Define the action of the Lie algebra of vector fields $h(s)\partial_s$,
$h\in {\mathbb C}[s,s^{-1}]$ as follows  
$$
h(s)\partial_s\Big(f(s){:}e_*^{(\tau^{-1}{+}s^2)(w_*^2{+}\nu)}{:}_{\tau}\Big)
=(h(s)\partial_sf(s)){:}e_*^{(\tau^{-1}{+}s^2)(w_*^2{+}\nu)}{:}_{\tau}
{+}h(s)(2s)f(s){:}(w_*^2{+}\nu){*}e_*^{(\tau^{-1}{+}s^2)(w_*^2{+}\nu)}{:}_{\tau}).
$$ 
For simplicity, we denote this 
\begin{equation}\label{derivation}
h(s)\partial_s\{(f(s)\}
=\{h(s)\partial_sf(s)\}
{+}{:}\{h(s)(2s)f(s)\}{*}(w_*^2{+}\nu){:}_{\tau}.
\end{equation}
For later use we denote these operations on the generators:
\begin{equation}\label{forextend}
\begin{aligned}
&\{s^m\}+\{s^n\}=\{s^m{+}s^n\}, \quad s^m\{s^n\}=\{s^ms^n\}\\
&s^{n{+}1}\partial_s\{s^m\}
   =\{ms^{n{+}m}\}
{+}{:}\{2s^{n{+}m{+}2}\}{*}(w_*^2{+}\nu){:}_{\tau}.
\end{aligned}
\end{equation}
We use the notation $\{x^0\}$, but we do not use the notation $\{1\}$.

$\{{\mathbb C}[s,s^{-1}]\}=
{\mathbb C}[s,s^{-1}]{:}e_*^{(\tau^{-1}{+}s^2)(w_*^2{+}\nu)}{:}_{\tau}$ 
is a ${\mathbb C}[s,s^{-1}]$-module, called often ``loop algebra'', 
on which the Lie algebra 
${\mathbb C}[s,s^{-1}]\partial_s$ acts naturally as derivations, where
a derivation means that 
$$
h(s)\partial_s (f(s)\{g(s)\})=(h(s)\partial_s f(s))\{g(s)\}{+}f(s)h(s)\partial_s\{g(s)\} 
$$

By defining $[V(s), f(s)]=V(s)f(s)$, and 
$[f(s),g(s)]=0$, the direct product space 
$\{{\mathbb C}[s,s^{-1}]\}\oplus {\mathbb C}[s,s^{-1}]\partial_s $
has a Lie algebra structure including $\{{\mathbb C}[s,s^{-1}]\}$
as a commutative Lie ideal.

\bigskip
We denote by $V_{\tau}$ the vector space spanned by 
$$
{\rm{Res}}_{s=0}f(s)\partial_s^k{:}e_*^{(\tau^{-1}{+}s^2)(w_*^2{+}\nu)}{:}_{\tau};
\quad k\in {\mathbb N}, \quad f(s)\in {\mathbb C}[s,s^{-1}].
$$
That is 
$V_{\tau}={\mathbb C}[\tau,\tau^{-1},\nu,w^2]e^{\frac{\nu}{\tau}}e^{-\frac{w^2}{\tau}}.$

\medskip
The essential part of residue calculus is 
\begin{equation}\label{point}
{\rm{Res}}_{s=0}(\partial_{s}h(s))=0,\quad 
\forall h \in V_{\tau}[[s,s^{-1}]].
\end{equation}

\medskip
From a basic viewpoint of the conformal field theory, a non-trivial 
residue give a violation of the additive structure around
$s=0$.  Namely, the integration by parts gives
$$
{\rm{Res}}_{s}\{(f'(s)g(s))\}=-{\rm{Res}}_{s}\{f(s)g'(s)\}{-}
{\rm{Res}}_{s}(f(s)(2s)g(s)
{:}e_*^{(\tau^{-1}{+}s^2)(w_*^2{+}\nu)}{*}(w_*^2{+}\nu){:}_{\tau}).
$$
Denote the second term by 
${\rm{Res}}_{s}({:}\{f(s)(2s)g(s)\}{*}(w_*^2{+}\nu){:}_{\tau})$, which
is a symmetric bilinear form. 
Using this we extend the usual commutative structure 
on the space $\{{\mathbb C}[s,s^{-1}]\}{\oplus}V_{\tau}$ to a 
noncommutative product  by defining 
$$
(\{f(s)\},a){\ctt}(\{g(s)\},b)
=(\{f(s){+}g(s)\},a{+}b{-}
{\rm{Res}}_{s=0}({:}\{f(s)sg(s)\}{*}(w_*^2{+}\nu){:}_{\tau})
{+}{\rm{Res}}_{s=0}\{f'(s)g(s)\}.
$$
This gives a noncommutative extension of the usual additive operation.
However, we regard this as an extension of commutative 
Lie algebra $\{{\mathbb C}[s,s^{-1}]\}{\oplus}V_{\tau}$ by the form:
$$
[(\{f(s)\}, a), (\{g(s)\},b)]=(0,{\rm{Res}}_{s}\{f(s)'g(s)\}{-}{\rm{Res}}_{s}\{g(s)'f(s)\})
$$
for the purpose to extend this to an algebra. 
We now make its universal enveloping algebra, but note here that 
the multiplicative structure is nothing to do with  
the original multiplicative structure of ${\mathbb C}[s,s^{-1}]$. 
For that purpose, we extend first the vector space $V_{\tau}$ to 
the commutative algebra $\tilde V_{\tau}$ generated by  $V_{\tau}$. 
$$
\tilde V_{\tau}={\mathbb C}[\tau,\tau^{-1},\nu,w^2]
e^{\mathbb N\frac{\nu}{\tau}}e^{-\mathbb N\frac{w^2}{\tau}}.
$$
We define next   
$$
\{f(s)\}{\btt}\{g(s)\}
=\{f(s)g(s)\}+{\rm{Res}}_{s=0}\{f'(s)g(s)\}{+}
{\rm{Res}}_{s=0}({:}\{f(s)sg(s)\}{*}(w_*^2{+}\nu){:}_{\tau})
$$
$$
\{f(s)\}{\btt}\big({:}\{g(s)\}{*}(w_*^2{+}\nu){:}_{\tau}\big)
={:}\{f(s)g(s)\}{*}(w_*^2{+}\nu){:}_{\tau}
{+}{\rm{Res}}_{s=0}({:}\{f(s)g(s)\}{*}(w_*^2{+}\nu){:}_{\tau}).
$$
$$
\big({:}\{f(s)\}{*}(w_*^2{+}\nu){:}_{\tau}\big){\btt}\{g(s)\}
={:}\{f(s)g(s)\}{*}(w_*^2{+}\nu){:}_{\tau}
{+}{\rm{Res}}_{s=0}({:}\{f(s)g(s)\}{*}(w_*^2{+}\nu){:}_{\tau}).
$$
Furthermore, we define 
$$
\begin{aligned}
\big({:}\{f(s)\}{*}&(w_*^2{+}\nu)^k{:}_{\tau}\big){\btt}
\big({:}\{g(s)\}{*}(w_*^2{+}\nu)^{\ell}{:}_{\tau}\big)\\
&={:}\{f(s)g(s)\}{*}(w_*^2{+}\nu)^{k{+}\ell}{:}_{\tau}
{+}{\rm{Res}}_{s=0}({:}\{f(s)g(s)\}{*}(w_*^2{+}\nu)^{k{+}\ell}{:}_{\tau}).
\end{aligned}
$$
$$
A{\btt}a=a{\btt}A;\quad A\in {\mathfrak A}_{\tau}, \quad a \in {\tilde V}_{\tau}.
$$
These define commutative product except the term 
${\rm{Res}}_{s=0}\{f'(s)g(s)\}$ on the first line. 
We call this the Heisenberg vertex algebra and denote this by 
${\mathfrak A}_{\tau}$. 

\bigskip
Recall the action \eqref{derivation} 
$h(s)\partial_s\{f\}=\{h(s)f'(s)\}{+}{:}\{h(s)(2s)f(s)\}{*}(w_*^2{+}\nu){:}_{\tau}.$
This forms an action of the Lie algebra  
${\mathbb C}[s,s^{-1}]\partial_s$, called the Witt algebra: That is,
it holds   
$$
h\partial_s(k\partial_s\{f\}){-}k\partial_s(h\partial_s\{f\})=[h\partial_s, k\partial_s]\{f\}
$$ 
where $[h\partial_s, k\partial_s]=(hk'{-}kh')\partial_s$. 

Next, we  extend this as a derivation of ${\mathfrak A}_{\tau}$.
Namely, we define 
$$
\begin{aligned}
h\partial_s(\{f\}&{\btt}\{g\})=(h\partial_s\{f\}){\btt}\{g\}{+}\{f\}{\btt}(h\partial_s\{g\})\\
&=\big(\{hf'\}{+}{:}\{h(2s)f\}{*}(w_*^2{+}\nu){:}_{\tau}\big){\btt}\{g\}
  {+}\{f\}{\btt}\big(\{hg'\}{+}{:}\{h(2s)g\}{*}(w_*^2{+}\nu){:}_{\tau}\big).
\end{aligned}
$$

As the residues such as ${\rm{Res}}_{s=0}\{f'(s)g(s)\}$ do not involve
the variable $s$, it looks at a glance that
$h\partial_s[\{f\},\{g\}]=0$, but the term $(w_*^2{+}\nu){:}_{\tau}$
can act on the residue part. Indeed, the action 
$h\partial_s[\{f\},\{g\}]$ is given  as follows:
$$
\begin{aligned}
h\partial_s(\{f\}{\btt}\{g\})
=&\big(\{hf'\}{+}{:}\{h(2s)f\}{*}(w_*^2{+}\nu){:}_{\tau}\big){\btt}\{g\}
  {+}\{f\}{\btt}\big(\{hg'\}{+}{:}\{h(2s)g\}{*}(w_*^2{+}\nu){:}_{\tau}\big)\\
=&
\{hf'g\}{+}{\rm{Res}}\{(hf')'g\}{+}{\rm{Res}}{:}\{shf'g\}{*}(w_*^2{+}\nu){:}_{\tau}\\
 &\qquad{+}{:}\{2shfg\}{*}(w_*^2{+}\nu){:}_{\tau}{+}{\rm{Res}}{:}\{2shfg\}{*}(w_*^2{+}\nu){:}_{\tau} \\
&+\{fhg'\}{+}{\rm{Res}}\{f'hg'\}{+}{\rm{Res}}{:}\{sfhg'\}{*}(w_*^2{+}\nu){:}_{\tau}\\
 &\qquad{+}{:}\{2sfhg\}{*}(w_*^2{+}\nu){:}_{\tau}{+}{\rm{Res}}{:}\{2sfhg\}{*}(w_*^2{+}\nu){:}_{\tau}
\end{aligned}
$$
Hence by using 
$$
{\rm{Res}}\{(hf')'g\}{+}{\rm{Res}}\{(hf'g'\}{+}{\rm{Res}}{:}\{2shf'g\}{*}(w_*^2{+}\nu){:}_{\tau}=0,
$$
exchanging $f$ and $g$ gives 
\begin{equation}\label{Wittact}
\begin{aligned}
h\partial_s([\{f\},\{g\}]_{\btt})=
{\rm{Res}}{:}\{2sh(f'g{-}fg')\}{*}(w_*^2{+}\nu){:}_{\tau}.
\end{aligned}
\end{equation}
Note that this term is caused by terms such as
${:}\{2shf\}{*}(w_*^2{+}\nu){:}_{\tau}$, hence \eqref{Wittact} must vanish, 
if one can eliminate these terms by a change of generators. 

\subsection{Central extension caused by singularity}

To make these clearer, we consider these on generators by setting $x_m=\{s^m\}$.
Consider now the Lie algebra 
$$
\mathfrak g=\{\sum_{n\in{\mathbb Z}}c_n x_n; c_n{\in}{\mathbb C};
[x_m,x_n]=(m{-}n)a_{m{+}n{-}1}(\tau^{-1},\nu, w)\},
$$ 
where $a_{m{+}n{-}1}(\tau^{-1},\nu, w)$ are Laurent coefficients.
Next, we make its universal enveloping algebra ${\mathfrak A}_{\tau}$ by 
extending the vector space $V_{\tau}$ to an algebra 
$\tilde V_{\tau}$ generated by  
$\{e^{\tau^{-1}\nu}
\frac{1}{\sqrt{-\tau}}e^{{-}\frac{1}{\tau}w^2}\}$
under ordinary commutative product. 
${\mathfrak A}_{\tau}$ is a noncommutative associative algebra generated by 
infinitely many generators $\{x_k; k\in\mathbb Z\}$ together with
commutation relations $[x_m,x_n]=(m{-}n)a_{m{+}n{-}1}(\tau^{-1},\nu, w)$.
In the case $\nu=0$ and $w=0$, we see that 
$$
[x_m,x_n]=2m\delta_{m+n,0}\frac{1}{\sqrt{-\tau}}, \quad
a_{m{+}n{-}1}(\tau^{-1},0, 0)=0, \quad a_{{-}1}(\tau^{-1}0, 0)=\frac{1}{\sqrt{-\tau}},  
$$ 
but in general 
$x_m{\btt}x_{n}=x_{n}{\btt}x_m{+}(m{-}n)a_{m{+}n{-}1}(\tau^{-1},\nu, w).$

\medskip
Let $E^{(k)}$ be the linear space spanned by 
$
{\tilde V}_{\tau}x_{n_1}{\btt}x_{n_2}{\btt}\cdots{\btt}x_{n_k}$.
It is not hard to see that the space $E^{(2)}$ consisting of 
all quadratic forms such as $\sum c_{mn}x_{m}{\btt}x_{n}$ form a 
Lie algebra acting on the space $E^{(1)}$ under the commutator 
bracket product $[a,b]_{\btt}=a{\btt}b{-}b{\btt}a$,
 i.e. $[E^{(2)}, E^{(1)}]=E^{(1)}$. 
This extends naturally on ${\mathfrak A}_{\tau}$ as derivations: i.e. 
$$
[E^{(2)},{\mathfrak A}_{\tau}]
\subset\tilde{\mathfrak A}_{\tau},
\quad [A, f{\btt}g]=[A,f]{\btt}g{+}f{\btt}[A,g]. 
$$

We want to write the extended action $h(s)\partial_s\{f(s)\}$ on the generators. 
 Recalling  
\begin{equation*}
s\partial_s\{s^m\}
=\{ms^{m}\}
{+}{:}\{2s^{m{+}2}\}{*}(w_*^2{+}\nu){:}_{\tau},
\end{equation*}
we define 
$$
[L_0, x_m]=mx_{m}{+}2{:}x_{m{+}2}{*}(w_*^2{+}\nu){:}_{\tau}.
$$
Since 
$$
[s^nL_{0},x_m]=s^n(mx_m{+}2{:}x_{m{+}2}{*}(w_*^2{+}\nu){:}_{\tau})
=mx_{n{+}m}{+}2{:}x_{n{+}m{+}2}{*}(w_*^2{+}\nu){:}_{\tau},
$$
we set $L_n=s^nL_0$ and define 
$$
[L_n,x_m]=mx_{n{+}m}{+}2{:}x_{n{+}m{+}2}{*}(w_*^2{+}\nu){:}_{\tau}
$$
Then, 
$$
\begin{aligned}
{}[L_\ell,[L_n,x_m]]
&=m(n{+}m)x_{n{+}m{+}\ell}{+}2(n{+}m{+}2){:}x_{n{+}m{+}\ell{+}2}{*}(w_*^2{+}\nu){:}_{\tau}\\
&\qquad +4(n{+}m{+}\ell{+}4){:}x_{n{+}m{+}\ell{+}4}{*}(w_*^2{+}\nu)_*^2{:}_{\tau}.
\end{aligned}
$$
It follows 
$$
{}[L_n,[L_\ell,x_m]]-[L_\ell,[L_n,x_m]]
=m(n{-}\ell)x_{n{+}m{+}\ell}{+}2(n{-}\ell)x_{n{+}m{+}\ell{+}2}{:}{*}(w_*^2{+}\nu){:}_{\tau}
$$
Thus, this is an action of the Witt algebra
$$
{}[[L_n,L_\ell],x_m]
=(n{-}\ell)[L_{n{+}\ell}, x_m ].
$$

The direct computation shows the following  
\begin{prop}\label{normalizae}
For every interger $m$, an element 
$y_m{=}
\sum_{k=0}^{\infty}\frac{(-2)^k}{k!}{:}x_{m{+}k}{*}(w_*^2{+}\nu)^k{:}_{\tau}$ 
written as a formal power series of ${:}(w_*^2{+}\nu)_*^k{:}_{\tau}$ 
satisfies $[L_0, y_m]=my_m$. It follows 
$$
[L_n, y_m]=s^n[L_0,y_m]=ms^ny_m=my_{n{+}m}.
$$
\end{prop}

Note that 
$$
y_m= \sum_{k=0}^{\infty}s^{m{+}k}\frac{(-2)^k}{k!}{:}x_{0}{*}(w_*^2{+}\nu)^k{:}_{\tau}
=s^m
\sum_{k=0}^{\infty}s^{k}\frac{(-2)^k}{k!}
{:}(w_*^2{+}\nu)^k{*}e_*^{(\tau^{-1}{+}s^2)(w_*^2{+}\nu)}{:}_{\tau}.
$$ 
Hence this is defined only as a formal power series in general, for this is 
$$s^m{:}e_*^{-2s(w_*^2{+}\nu)}{*}e_*^{(\tau^{-1}{+}s^2)(w_*^2{+}\nu)}{:}_{\tau}=
s^me_*^{(\tau^{-1}{+}s^2{-}2s)(w_*^2{+}\nu)}{:}_{\tau}
$$
and this diverges at $s=2$. 
Although the expression seems a slightly confusing, it is convenient to
view this 
$$
y_m={:}e_*^{-2s(w_*^2{+}\nu)}s^m{*}e_*^{(\tau^{-1}{+}s^2)(w_*^2{+}\nu)}{:}_{\tau}
={:}e_*^{-2s(w_*^2{+}\nu)}{*}x_m{:}_{\tau}
$$  
as it is inverted easily by 
$$
{:}y_m{*}e_*^{2s(w_*^2{+}\nu)}{:}_{\tau}=s^m{:}e_*^{(\tau^{-1}{+}s^2)(w_*^2{+}\nu)}{:}_{\tau}=x_m.
$$

\subsubsection{Heisenberg vertex algebra}

As it is easy to see  
$$
[{:}x_{m}{*}(w_*^2{+}\nu)^k{:}_{\tau},{:}x_{n}{*}(w_*^2{+}\nu)^\ell{:}_{\tau}]
=[x_m,x_n]{*_{\tau}}{:}(w_*^2{+}\nu)^{k{+}\ell}{:}_{\tau},
$$
the commutator $[y_m,y_n]$ belongs to the space of formal power series  
${\tilde V}_{\tau}{*_{\tau}}[[{:}(w_*^2{+}\nu){:}_{\tau}]]$. This is
the space of all formal power series written in the form 
$$
\sum_k a_k(\tau^{-1},\nu, w){*_{\tau}}{:}(w_*^2{+}\nu)_*^k{:}_{\tau},
\quad a_k(\tau^{-1},\nu, w)\in {\tilde V}_{\tau}. 
$$

\medskip
Set $[y_m,y_n]=C_{m,n}$. By Proposition\,\ref{normalizae} and by the
remark below \eqref{Wittact}, we see  
$[L_n,C_{m,n}]=0$. 
Since $L_k$ acts as a derivation, Jacobi 
identity of Lie algebra gives 
restriction to the constants $C_{m,n}$:
$$
0=[L_k,[y_{\ell},y_m]]=[[L_k,y_{\ell}],y_m]{+}[y_{\ell},[L_{k},y_m]]=
\ell[y_{\ell{+}k},y_m]{+}m[y_{\ell},y_{m{+}k}].
$$
Hence 
\begin{equation}\label{equal}
\ell C_{\ell{+}k,m}{+}mC_{\ell, m{+}k}=0.
\end{equation}
Set $k=0$ to obtain $C_{\ell,m}=c_m\delta_{\ell{+}m,0}$. Set $m=1$ 
further in \eqref{equal} to obtain  
$\ell c_1\delta_{\ell{+}k{+}1,0}{+}c_{k{+}1}\delta_{\ell{+}k{+}1,0}=0.$ 
Hence, we have $c_m=mc_1$, 
$$
\begin{aligned}
c_1=&C_{-1,1}=(-2)\sum_{k,\ell}\frac{(-2)^{k{+}\ell}}{k!\ell !}
a_{k{+}\ell-1}(\nu,\tau^{-1},w){:}(w_*^2{+}\nu)^{k{+}\ell}{:}_{\tau}\\
=&(-2)\sum_{n}\frac{4^{2n}}{(2n)!}
{:}a_{2n{-}1}(\nu,\tau^{-1},w){*}(w_*^2{+}\nu)_*^{2n}{:}_{\tau}.
\end{aligned}
$$
Consequently, we have 
\begin{prop} 
The system $\{y_m, m\in{\mathbb Z}\}$ has the property that , 
$$
\begin{aligned}
&[y_m, y_n] = m\delta_{m+n,0}c_1,\quad c_1\in {\tilde V}_{\tau}{*_{\tau}}[[{:}(w_*^2{+}\nu){:}_{\tau}]]\\
&[L_m, y_n]=my_{m{+}n}\\
&[[L_m,L_n],y_\ell]=[(m{-}n)L_{m{+}n},y_{\ell}] 
\end{aligned}
$$
It follows in particular $[y_0, y_m]=0$ for every $y_m$.
In particular, as there is no zero-divisor in 
${\tilde V}_{\tau}$, 
if $\sum_k a_ky_k$, $a_k\in {\tilde V}_{\tau}$, satisfies 
$[\sum_k a_ky_k, y_m]=0$ for every $y_m$, then $\sum_k a_ky_k=a_0y_0$. 
\end{prop}

$\{y_m; m\in{\mathbb Z}\}$ forms a standard basis of Heisenberg vertex algebra over 
${\tilde V}_{\tau}{*_{\tau}}[[{:}(w_*^2{+}\nu){:}_{\tau}]]$. 

\bigskip
So far, $L_m$ is not an established element defined only 
as an adjoint operator $[L_m,{\cdot }]$ acting on $E^{(1)}$. 
The following theorem is known as Sugawara construction:
\begin{thm}\label{Sugawara}
Elements of Witt algebra can be represented by elements of $E^{(2)}$. 
\end{thm}

Thus, regarding $L_m$ as an element of $E^{(2)}$, we set 
$[L_m,L_n]=(m-n)L_{m{+}n}+K_{m,n}$, as 
$$
[K_{m,n},y_{\ell}]=
[L_m,[L_n,y_{\ell}]]-[L_n,[L_m,y_{\ell}]]-(m-n)[L_{m{+}n}, y_{\ell}]=0.
$$
$K_{m,n}$ must be central elements. Such a central extension of 
Witt algebra is called the Virasoro algebra.

Such an  extended Lie algebra is known to be isomorphic to the one defined by 
\begin{equation}\label{virasoro22}
[L_m,L_n]=(m{-}n)L_{m{+}n}{+}
c(\nu,\tau^{-1},w)\frac{m(m^2{-}1)}{12}\delta_{m{+}n,0},
\quad [L_n,c(\nu,\tau^{-1},w)]=0.
\end{equation} 

Note that restricting $n$ even integers $\{L_{2n}; n{\in}{\mathbb Z}\}$
forms a Lie subalgebra.

\end{document}